\documentclass[11pt,a4paper]{article}
\usepackage{jheppub}
\usepackage[utf8]{inputenc} 
\usepackage{mathtools,xcolor,derivative,braket,cancel}
\hypersetup{citecolor=purple}

\newcommand{\UV}{{\textrm{UV}}}

\newcommand{\eff}{{\textrm{eff}}}
\newcommand{\SUSYbrk}{{\cancel{\textrm{SUSY}}}}
\newcommand{\pl}{{\textrm{Pl}}}
\newcommand{\hc}{{\textrm{h.c.}}}
\renewcommand{\O}{{\mathcal{O}}}
\newcommand{\N}{{\mathcal{N}}}
\renewcommand{\L}{{\mathcal{L}}}
\newcommand{\R}{{\mathcal{R}}}
\newcommand{\D}{{\mathcal{D}}}
\newcommand{\bs}[1]{\boldsymbol{#1}}
\newcommand{\DM}{{\textrm{DM}}}
\newcommand{\osc}{{\textrm{osc}}}
\newcommand{\anh}{{\textrm{anh}}}

\def\shrinkage{2.1mu}
\def\vecsign{\mathchar"017E}
\def\dvecsign{\smash{\stackon[-2.13pt]{\mkern-\shrinkage\vecsign}{\rotatebox{180}{$\mkern-\shrinkage\vecsign$}}}}
\def\dvec#1{\def\useanchorwidth{T}\stackon[-4.88pt]{#1}{\,\dvecsign}}
\usepackage{stackengine}
\stackMath

\def\bea  {\begin{eqnarray}}   \def\eea  {\end{eqnarray}}
\def\TeV{\,{\rm TeV}}
\def\GeV{\,{\rm GeV}}

\def\keV{\,{\rm keV}}
\def\eV{\,{\rm eV}}

\def\chimin{\chi_{\rm min}}

\begin{document}
	
	\title{Ultralight Dilatonic Dark Matter}
	
	\author[a]{Abhishek Banerjee,}
	\author[b]{Csaba Cs\'aki,}
	\author[c]{Michael Geller,}
	\author[c]{Zamir Heller-Algazi}
	\author[d]{and Ameen Ismail}
	
	\affiliation[a]{Maryland Center for Fundamental Physics, University of Maryland, College Park, MD 20742, USA}
	\affiliation[b]{Department of Physics, LEPP, Cornell University, Ithaca, NY 14853, USA}
	\affiliation[c]{School of Physics and Astronomy, Tel-Aviv University, Tel-Aviv 69978, Israel}
	\affiliation[d]{Enrico Fermi Institute, University of Chicago, Chicago, IL 60637, USA}
	
	\emailAdd{abanerj4@umd.edu}
	\emailAdd{csaki@cornell.edu}
	\emailAdd{mic.geller@gmail.com}
	\emailAdd{zamir.heller@gmail.com}
	\emailAdd{ameenismail@uchicago.edu}
	
	\abstract{
    The dilaton, a pseudo-Nambu-Goldstone boson (pNGB) of broken scale invariance, is an appealing ultralight dark matter (DM) candidate. Its mass is protected by conformal invariance and it can be searched for in tabletop experiments. However, contrary to standard pNGBs of internal symmetries, the dilaton generically has a large non-derivative self-coupling, leading to radiative contributions to its mass of the order of its decay constant. Hence typical ultralight dilatons should also have sub-eV decay constants, which would incur significant deviations from standard DM behavior at structure formation times, in severe tension with observations. Therefore, a fine-tuning is required to generate a hierarchy between the mass and the decay constant. In this work, we consider whether supersymmetry (SUSY) can be used to protect this hierarchy from quantum corrections. To ensure an ultralight dilaton mass robust against realistic SUSY-breaking contributions, we must consider a novel dilaton stabilization mechanism. The observed DM abundance can be produced by the misalignment mechanism for dilaton masses ranging from $10^{-11}$ to $1$ eV. Unfortunately, irreducible SUSY-breaking corrections due to gravity restrict the couplings between the dilaton and the Standard Model to be extremely small, beyond the reach of any current or proposed experiments. Our work demonstrates that constructing a consistent model of ultralight dilaton DM is quite involved.
    }

    \arxivnumber{2506.21659}
	
	\maketitle
	\flushbottom
	
	\section{Introduction}

    Most of the matter in the Universe consists of dark matter (DM), a cold and collisionless fluid. So far, it has only been observed indirectly through its gravitational interactions with the Standard Model (SM) particles; its microscopic nature, including whether it has any nongravitational interactions with the SM, remains unknown. Among the many possible DM models, ultralight dark matter (ULDM) --- DM lighter than $O(\!\eV)$ --- has attracted a lot of attention in recent years. This is because the number density of ULDM is large, such that it behaves as a classical field, and its wave nature enables new detection strategies as well as rich astrophysical and cosmological phenomena; see~\cite{Jaeckel:2022kwg, Hui:2016ltb,Hui:2021tkt} for reviews.
    
    Theoretical models of ultralight dark matter should be able to explain two main properties. The first is the smallness of the ULDM mass, which is generically sensitive to radiative corrections that could spoil its lightness. The second is the observed DM relic abundance, which must be generated by a nonthermal production mechanism since the mass is below the warm DM bound. The axion/axion-like particle (ALP)~\cite{Marsh:2015xka} and the dark photon~\cite{Antypas:2022asj,Broadberry:2024yjw} are the only well-established ULDM candidates with both of these properties.

    Alternatively, one can consider a generic scalar as ULDM, with a feeble linear coupling to the SM. However, any sizable couplings between it and the SM would generate large radiative corrections to its mass. Thus, one is forced to either fine-tune the ULDM mass, or to consider very feeble couplings that are difficult to probe experimentally. To bypass this issue, theories where the ULDM is a pseudo-Nambu-Goldstone boson (pNGB) have been proposed, where an approximate symmetry protects the mass whilst allowing for large couplings to the SM; see~\cite{Piazza:2010ye,Banerjee:2018xmn,Brzeminski:2020uhm} for studies following this direction.
    
    An appealing proposal is to consider the case where the ULDM is a dilaton, the pNGB of spontaneously broken scale invariance~\cite{Antypas:2022asj,Cho:1998js,Damour:2010rm,Damour:2010rp,Arvanitaki:2014faa}. The scaling symmetry protects the mass of the dilaton because it universally couples to the trace of the stress-energy tensor, resulting in a tiny mass correction of order $H_0\sim 10^{-33}\eV$.
    However, unlike pNGBs of internal symmetries (e.g. the QCD axion), the mass of the dilaton $m_\chi$ is generically of the same order (up to a loop factor) as the decay constant $f$, the scale of spontaneous symmetry breaking (SSB)~\cite{Randall:1999ee,Bellazzini:2014yua,Panico:2015jxa,Csaki:2004ay,Csaki:2018muy}. If the dilaton is DM, the lack of separation between the mass and the decay constant leads to tension with structure formation in the early Universe~\cite{Hubisz:2024hyz}. At the onset of structure formation, when the temperature of the SM is $T\sim \keV$, the observed DM energy density is $\rho_\DM\approx T^3\cdot \eV$~\cite{Kuhlen:2012ft}. At this time, the dilaton oscillates about its minimum with energy density $\rho_\chi\approx m_\chi^2\delta\chi^2$ and redshifts like cold DM. By matching $\rho_{\rm DM}$ with $\rho_\chi$, we see that the amplitude of the field oscillations must be much larger than the dilaton mass, $\delta\chi\gg m_\chi$. But the dilaton only redshifts as cold DM if $\delta\chi\lesssim f$; otherwise, the potential is not quadratic and the oscillations are anharmonic. Therefore, generic ultralight dilaton DM is inconsistent with cosmological observations. One needs to separate the dilaton mass $m_\chi$ and the decay constant $f$ to make the dilaton a viable ULDM candidate.\footnote{Because the dilaton mass is suppressed by powers of the anomalous dimension $\epsilon$, it is possible to get a light dilaton if the anomalous dimension is small~\cite{Coradeschi:2013gda,Bellazzini:2013fga,Girmohanta:2024ywz}.
    However, if the anomalous dimension is too small then the UV scale is trans-Planckian, due to its exponential sensitivity to $1/\epsilon$~\cite{Hubisz:2024hyz}.}

    The goal of this work is to construct a natural, UV-complete model of ultralight dilaton DM.
    We invoke supersymmetry (SUSY) to protect a hierarchy between the dilaton mass and the SSB scale. In a non-SUSY theory, radiative corrections generate an unavoidable $O(1)$ quartic term in the dilaton effective potential, which leads to a large dilaton mass. In our SUSY model, we can set the quartic to be small and it remains small, as ensured by the nonrenormalization theorem. Hence, SUSY allows us to have a parametrically light dilaton that is a suitable ultralight DM candidate. The effective SUSY-breaking scale can be much lower in the dark sector than in the SM if SUSY is broken at a low scale, and communicated to the visible sector via gauge mediation but to the dark sector via gravity mediation.

    Specifically, the UV model of the dark sector (DS) is a supersymmetric CFT (SCFT) where the dilaton DM arises as the pNGB of spontaneously broken scale invariance. The IR mass gap is triggered by an external operator that explicitly breaks scale invariance in the UV.
    The dilaton mass is $m_\chi = \lambda f$, where $\lambda \ll 1$ is a small parameter proportional to the VEV this operator obtains in the IR. The holographic dual is a SUSY Randall-Sundrum (RS) model, where the dilaton DM is identified with the radion field that sets the size of the extra dimension. The operator that generates the mass gap is dual to a field in the bulk, and the small parameter $\lambda$ arises as the tadpole of the bulk field on the IR brane.

    The major model-building challenge is that realistic SUSY-breaking corrections --- with a SUSY-breaking scale consistent with current LHC data --- generically make the dilaton mass too large to be ultralight DM. In the Goldberger-Wise mechanism~\cite{Goldberger:1999uk,Goldberger:1999un} (as well as the relevant stabilization mechanism~\cite{Csaki:2023pwy}) the SUSY-breaking terms must be sufficiently small to not affect the dilaton stabilization, which generically results in the bound $m_\chi\gtrsim \sqrt F$, where $\sqrt F$ is the effective SUSY-breaking scale in the dark sector. We overcome this issue by employing a somewhat unusual stabilization mechanism: we trigger the spontaneous breaking of scale invariance with an irrelevant external operator induced by SUSY-breaking. By tying the breaking of the two symmetries to each other, we ensure that $m_\chi \propto \sqrt F$, and because the operator is irrelevant its effects in the IR are further warped down, allowing for $m_\chi \ll \sqrt F$. In the 5D dual, $F$ is the VEV of the auxiliary component of the bulk field dual to the external operator. With our stabilization mechanism, the dilaton can remain ultralight even after accounting for realistic SUSY-breaking corrections.

    We will show that the dilaton ULDM can be produced by the misalignment mechanism, although the details differ from ALP production. These differences arise because the dilaton field is noncompact and its potential is not periodic, which are important when the dilaton is initially far from the potential minimum. In this case the dilaton undergoes an epoch of anharmonic oscillations, during which its energy density redshifts faster than radiation, before settling into coherent harmonic oscillations about the minimum. In addition, we find that because the dilaton couples to the energy density of the Universe, during inflation it redshifts towards zero. This limits the possible initial displacement from the vacuum, unless one considers modifications to the inflaton-dilaton couplings. The final DM relic abundance is greatly affected by these differences. 

    To the best of our knowledge, our construction is the first UV-complete model of ultralight dilaton DM with a consistent cosmological history. Unfortunately, we find that anomaly-mediated SUSY-breaking (AMSB) corrections generate unacceptably large contributions to the dilaton mass unless its coupling to the SM is very small. Furthermore, unless there are nonminimal couplings between the dilaton and the inflaton, matching the correct abundance predicts even tinier dilaton interactions with the SM. Thus, contrary to our initial hope, we find that dilaton ULDM requires couplings to the SM that are small and well beyond the reach of proposed experiments, similar to generic scalar ULDM. Although ultralight dilaton DM is frequently considered in phenomenological studies, our work demonstrates that a realistic theory is difficult to realize and requires rather baroque ingredients.

    The paper is organized as follows. In Sec.~\ref{sec_low_energy_summary} we give a brief summary of the main low energy physics of our model. We explain the UV construction of our model in Sec.~\ref{sec_SCFT}, where we calculate the dilaton mass and show that it is hierarchically below the IR scale. We compute the abundance of ultralight dilaton DM in Sec.~\ref{sec_misalignement}, with a focus on the unique anharmonic misalignment production scenario. In Sec.~\ref{sec_phenomenology} we derive the ULDM dilaton phenomenology and plot the available parameter space of our model. We find that ULDM dilaton can comprise all of DM for masses between $10^{-11}$--$1$\eV, and currently no existing or planned experiments are probing the relevant region of parameter space.

    \section{Summary of the Low-Energy Physics}\label{sec_low_energy_summary}

    The dark sector of our model consists of two fields: the scalar dilaton $\chi$ and the pseudoscalar R-axion $\vartheta$. They are both singlets under the SM gauge group with almost identical masses $m_\chi\approx m_\vartheta$. The dilaton and the R-axion make up the entirety of DM, and are produced by the misalignment mechanism, as will be explained in Sec.~\ref{sec_misalignement}. The dark sector also contains a light dilatino, the fermionic superpartner of the dilaton and the R-axion. However, it is very weakly coupled and rarely produced in the early Universe, so it plays no role in cosmology and is irrelevant to experiments.
    
    In Sec.~\ref{sec_SCFT} we present the UV model for the dark sector, and show it results in  a scalar potential of the form 
    \begin{equation}
        V_\eff = \frac{\frac{1}{2}m_\chi^2f^2}{\epsilon-1}\left[\frac{1}{1+\epsilon}\left(\frac{\overline\chi}{f}\right)^{2+2\epsilon}-\frac{2}{3+\epsilon}\left(\frac{\overline\chi}{f}\right)^{3+\epsilon}\cos[(3+\epsilon)\overline\vartheta/f]\right].
    \end{equation}
    Here $\overline\chi$ and $\overline\vartheta$ are respectively the (canonically normalized) dilaton and R-axion, $f$ is the minimum of the dilaton potential corresponding to the scale of SSB, and $\epsilon$ is dimensionless and greater than $1$. The dilaton mass and its VEV are independent of each other, and we are free to assume that $m_\chi\ll f$.

    The most important interaction of the dark sector with the SM is through the dilaton portal, as we calculate in Sec.~\ref{sec_phenomenology}. The dilaton couples linearly to the trace of the stress-energy tensor, resulting in couplings to the electron and the photon given by
    \begin{equation}
		\L_{\chi-\rm{SM}}=\frac{\overline\chi}{\Lambda} T^\mu_\mu \supset \frac{\overline\chi}{\Lambda} \left(m_e \overline{e} e + \frac{\beta(e)}{2e} F_{\mu\nu}^2 \right)
	\end{equation}
	where $m_e$ is the electron mass, $\beta(e)$ is the QED beta function and $\Lambda=6M_\pl^2/f$ is the suppression scale, which is trans-Planckian. After we work out the production mechanism and the phenomenology, we present exclusion plots in Figs.~\ref{fig:exclusion2} and~\ref{fig:exclusion}, assuming two different initial conditions for the dilaton. 
	
	\section{UV Model}\label{sec_SCFT}
	
	\subsection{Setup}\label{sec_setup}

\begin{figure}
		\centering
        \includegraphics[width=0.7\textwidth]{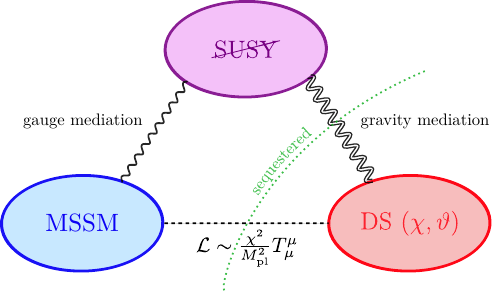}
		\caption{A sketch of our model. The DS is a SCFT where the role of DM is played by the dilaton $\chi$ and the R-axion $\vartheta$. The SUSY-breaking sector transmits SUSY-breaking to the MSSM via gauge mediation and to the DS via gravity mediation. The DS is sequestered from the other sectors, and interacts with the MSSM via the coupling of the dilaton to the trace of the stress-energy tensor.}
		\label{fig_sector_schematic}
	\end{figure}

    There are three sectors in our UV model: the first two are the visible sector and the dark sector. Since we are assuming an underlying SUSY, the third sector is the standard SUSY-breaking sector, which in turn is mediated to the first two sectors. For the dark sector we are interested in DM with masses below $1\eV$ which, as we shall see, requires a very low SUSY-breaking scale. Therefore we will assume SUSY-breaking is communicated by gravity mediation to the dark sector and also that the ultimate SUSY-breaking scale is low, $M_\SUSYbrk\sim O(100)\TeV$. In the visible sector, which contains the MSSM, SUSY must be broken above $O(10)\TeV$ for the superpartners to be heavy enough to avoid LHC limits~\cite{ATLAS:2024lda}. We therefore assume that SUSY-breaking is gauge-mediated to the MSSM where its effects are suppressed by a loop factor (see~\cite{Giudice:1998bp} for a review). Our construction\footnote{Ref.~\cite{Cacciapaglia:2023syp} recently proposed the same setup for heavier DM \textit{above} the effective SUSY-breaking scale in the DS, while we consider ultralight DM \textit{below} it. SUSY-breaking effects are very important in our scenario, while in~\cite{Cacciapaglia:2023syp} they are negligible.} is illustrated in Fig.~\ref{fig_sector_schematic}. 
    
\begin{figure}
		\centering
        \includegraphics[width=0.55\textwidth]{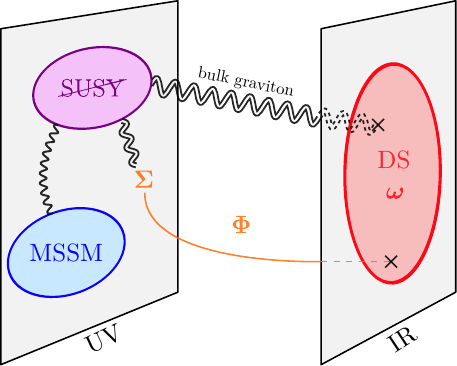}
		\caption{A sketch of our model in a 5D braneworld. The DS is the modulus field $\bs\omega$, whose radial component is the dilaton which determines the size of the extra dimension. The SUSY-breaking and MSSM sectors are localized on the UV brane, and a bulk field $\bs\Phi$ gets a VEV $\bs\Sigma$ on the UV brane. $\bs\Phi$ stabilizes the size of the extra dimension when the auxiliary component of $\bs\Sigma$ receives a VEV from SUSY-breaking via gravity mediation. SUSY-breaking is transmitted to the MSSM via gauge mediation, and to the DS on the UV brane via gravitons propagating through the bulk.}
		\label{fig_5D_sector_schematic}
	\end{figure}

    We model the dark sector as a strongly-coupled, spontaneously broken CFT near a fixed point, or equivalently as a warped 5D Randall-Sundrum model~\cite{Randall:1999ee} that is holographically dual to the CFT~\cite{Rattazzi:2000hs,Arkani-Hamed:2000ijo}. The MSSM and the SUSY-breaking sectors are elementary and the dark sector is composite. In the 5D picture, the MSSM and the SUSY-breaking sectors are UV-localized and the dark sector is IR-localized, see Fig.~\ref{fig_5D_sector_schematic}. Crucially, the CFT/RS is \textit{sequestered} from the other two sectors~\cite{Randall:1998uk,Luty:2001jh,Luty:2001zv,Schmaltz:2006qs}; that is, the external operators that explicitly break conformal invariance are all irrelevant (UV-localized in 5D).
    The sequestering suppresses dangerous local SUSY-breaking operators that would induce mass corrections that are too large for ultralight DM. Indeed, the SUSY-breaking scale in the dark sector is $m_{3/2}\approx M_\SUSYbrk^2/M_\pl\gtrsim10\eV$ due to gravity-mediation (we identify this scale as the gravitino mass). Therefore, Planck-suppressed contact terms with the SUSY-breaking sector would generically induce $O(m_{3/2})\gtrsim10\eV$ mass corrections that are unacceptably large for ULDM.

    The ultralight DM in our model is the dilaton of the CFT, dual to the radion of the RS model. The dilaton is a natural ULDM candidate because its mass is protected by the scale invariance of the CFT. In Secs.~\ref{sec_SCFT_dilaton_stabilization} and~\ref{sec_spectrum} we will use the gravity-mediated SUSY-breaking in the UV to explicitly break the scale invariance of the CFT and generate a mass gap, such that the dilaton acquires a mass proportional to $m_{3/2}$, which vanishes when SUSY is exact. In the 5D description, the radion VEV is arbitrary and thus the radion is massless, and it is stabilized by a massive bulk field that obtains a VEV~\cite{Goldberger:1999uk,Goldberger:1999un}. As will be explained in App.~\ref{app_5D_RS_EFT_proof}, SUSY-breaking on the UV brane generates the VEV of the bulk field responsible for stabilizing the IR brane, similar to the CFT picture.
	
	\subsection{Dilaton Stabilization}\label{sec_SCFT_dilaton_stabilization}

    The dilaton of the CFT plays the role of ultralight DM in our model. A cosmological history consistent with observations requires the IR scale at which the CFT is spontaneously broken to be hierarchically larger than the dilaton mass. Standard dilaton stabilization mechanisms do not separate these two scales, so we require a different stabilization mechanism which we describe below.
	
	We are analyzing a strongly-coupled SCFT that is spontaneously broken in the IR. The spontaneous breaking of scale invariance generates an NGB mode $\chi$ which is the dilaton field, and its VEV sets the IR scale where the CFT is spontaneously broken. We wish to generate a large hierarchy between the UV cutoff and the IR scale, so the dilaton must obtain a nonzero VEV exponentially below the UV cutoff. Due to the strong coupling, the dilaton potential cannot be calculated directly from the CFT, but we can determine the form of the effective dilaton potential $V_\eff(\chi)$ using spurion analysis. Scale-invariant quartic $\chi^4$ terms in $V_\eff(\chi)$ either push $\braket{\chi}$ to 0 (conformal invariance is unbroken) or toward the UV cutoff, and if $V_\eff(\chi)=0$ the dilaton is massless and the IR scale is arbitrary. In order to stabilize the dilaton about $\braket{\chi}\neq0$ and give it a mass, we need to introduce an explicit source of conformal symmetry breaking. This explicit breaking should be small enough that scale invariance is mostly broken spontaneously and the dilaton description is still valid at low energies. 
	
	We perturb the CFT in the UV by turning on an operator $\delta\L=g\O$ with a small coupling $g$. Since $g$ is small in the UV, the CFT remains nearly conformal until the coupling $g$ reaches a critical value where the operator $\O$ obtains a nonzero expectation value, triggering the spontaneous breaking of conformal invariance. The effective dilaton potential is found by spurion analysis: $g$ runs as $g(\mu)=g(\mu_\UV)(\mu_\UV/\mu)^{-\epsilon}$, where $\epsilon$ is the anomalous dimension of $\O$, so we can restore scale invariance by treating $g$ as having scaling dimension $[g]=-\epsilon$. This corresponds to polynomial terms $P[g(\chi)]\chi^4\supset \chi^{4+\epsilon}\mu_\UV^{-\epsilon},\chi^{4+2\epsilon}\mu_\UV^{-2\epsilon},\ldots$ in the dilaton potential that stabilize its VEV at a nonzero value. The size of $g$ sets the VEV of the dilaton $\braket{\chi}$ and thus the IR scale, and the dilaton is now a pNGB whose mass is related to the source of explicit breaking $\O$.
	
	In our model we are considering a \textit{super}-conformal field theory, so the argument above needs to be augmented to incorporate SUSY. In $\N=1$ SUSY this can be done by working in superspace. When an SCFT is spontaneously broken, in addition to the dilaton NGB $\chi$ associated with broken scale invariance there is an ``R-axion'' NGB $\vartheta$ associated with broken superconformal $R$-symmetry~\cite{Nelson:1993nf}. The dilaton is then promoted to a chiral superfield $\bs\omega=\{\omega,\tilde\omega,F_\omega\}$, which we call the ``modulus'' superfield~\cite{Luty:2000ec}. Its scalar component is $\omega\equiv\chi e^{i\vartheta}$ and its fermion partner $\tilde\omega$ is the ``dilatino''. The low-energy theory is again determined by spurion analysis, and an explicit source of conformal symmetry breaking is required to make the modulus massive with a nontrivial VEV. In addition to the requirement that this explicit breaking be small so that the theory remains nearly conformal, the SUSY-breaking in the UV must also remain small so that the theory remains nearly supersymmetric down to the IR scale.
	
	We will trigger the breaking of conformal invariance in the SCFT with a chiral operator $\bs\O=\{\O,\tilde\O,F_{\O}\}$ with scaling dimension $[\O]=3+\epsilon$, where $\epsilon$ is identified as the anomalous dimension of $F_\O$.\footnote{The fermion and auxiliary components of $\bs\O$ have scaling dimensions $[ \tilde\O ]=[\O]+\frac{1}{2}$ and $[F_\O]=[\O]+1$.} $\bs\O$ weakly deforms the SCFT in the UV as
	\begin{equation}\label{eq_CFT_weak_SCFT_deformation}
		\delta\L=\int\odif[order=2]{\theta}\bs\Sigma\bs\O+\hc=\Sigma F_\O-\tilde\Sigma\tilde\O+F_\Sigma\O+\hc
	\end{equation}
	$\bs\Sigma=\{\Sigma,\tilde\Sigma\,F_\Sigma\}$ contains the couplings for each component of $\bs\O$ in the UV, and is often referred to in the literature as the ``source field'' (see e.g.~\cite{Cacciapaglia:2008bi}). The coupling $F_\Sigma$ represents sources of SUSY-breaking in the UV, and similarly the fermion coupling $\tilde\Sigma$ will mix with the goldstino $\tilde G$, the Nambu-Goldstone mode associated with SUSY breaking. $\bs\Sigma$ is the conformal symmetry breaking spurion in the effective low-energy theory, and is assigned a spurious scaling dimension $[\Sigma]=-\epsilon$ such that $\bs\Sigma\bs\omega^\epsilon$ terms are formally scale invariant.
	
	In App.~\ref{app_5D_RS_EFT_proof} we use the dual holographic description of the SCFT in a warped background~\cite{Rattazzi:2000hs} to find the correct form of the low-energy modulus EFT with SUSY taken into account. We denote $k$ as the UV scale $\mu_\UV$, and for convenience we rescale $\bs\Sigma\rightarrow\bs\Sigma k^{-\epsilon}$ so that the spurion field is dimensionless. In our convention, $\bs\Sigma$ is the usual dim-1 spurion divided by the mass of the mediator field $M_{\rm mess}$ so that $F_\Sigma=M_\SUSYbrk^2/M_{\rm mess}$. The holographic calculation in the AdS background gives the effective modulus Lagrangian
	\begin{equation}\label{eq_CFT_low_energy_SUSY_modulus_EFT}
    \begin{aligned}
        \L_\eff&=\frac{3N^2}{4\pi^2}\int\odif[order=4]{\theta}\bs\omega^\dagger\bs\omega+\int\odif[order=2]{\theta}\kappa\bs\omega^3+\hc\\
        &-\frac{1}{1+\epsilon}\int\odif[order=4]{\theta}\bs\omega^\dagger\bs\omega\left|\bs\Sigma\left(\frac{\bs\omega}{k}\right)^\epsilon\right|^2+\int\odif[order=2]{\theta}\lambda\bs\omega^3\bs\Sigma\left(\frac{\bs\omega}{k}\right)^\epsilon+\hc
    \end{aligned}
	\end{equation}
    The first line of this equation is scale invariant, where $N$ is the number of colors in the SCFT and $|\kappa|^2|\omega|^4$ is the quartic. It is evident that no spontaneous symmetry breaking can occur when $\kappa\neq0$, which admits a minimum at $\omega_{\rm min}=0$. Therefore, it is necessary to introduce a small explicit breaking in Eq.~\eqref{eq_CFT_weak_SCFT_deformation} that generates the second line in Eq.~\eqref{eq_CFT_low_energy_SUSY_modulus_EFT} involving the spurion $\bs\Sigma$. This nearly supersymmetric description assumes that the explicit SUSY-breaking spurion $F_\Sigma$ is small, which is easily achieved if we assume that SUSY-breaking is mediated to the CFT sector by gravity so that $|F_\Sigma|=m_{3/2}$ ($M_{\rm mess}=M_\pl$). We also assume that $\epsilon>1$ so that the operator $\O$ that couples to $F_\Sigma$ in the UV is irrelevant. A similar form of the superpotential was previously found in~\cite{Sundrum:2009gv} using spurion analysis, but it was missing the K\"ahler correction in the second line, which plays a crucial role in our model.

    Normally, the dilaton quartic $\chi^4$ in CFTs is an order-one number because it receives quantum corrections, so it cannot be small without tuning. In SUSY, the quartic $\kappa$ is exact due to the superpotential nonrenormalization theorem, so if it is set to be small or vanishing, it remains so. This allows the novel possibility of setting $\kappa=0$ in the SCFT, and consequently having \textit{no quartic dilaton interaction in the effective potential.} We will also assume that $\lambda$ is extremely small for the same reason, which will be the source of the UV/IR hierarchy. In contrast, the coefficient of the noncanonical term in the K\"ahler potential cannot be chosen and is fixed.

    We find two main contributions to the modulus scalar potential in Eq.~\eqref{eq_CFT_low_energy_SUSY_modulus_EFT}: the first is obtained directly from the nonzero VEV of $F_\Sigma$, 
    \begin{equation}\label{eq_CFT_SUSYbrk_modulus_potential}
		V_{F_\Sigma}=\frac{1}{1+\epsilon}|F_\Sigma|^2|\omega|^{2+2\epsilon}k^{-2\epsilon}-\lambda F_\Sigma \omega^{3+\epsilon} k^{-\epsilon} + \hc,
	\end{equation}
    and the second is obtained by solving for $F_\omega$,
	\begin{equation}\label{eq_CFT_SUSYcnsrv_modulus_potential}
      V_{F_\omega}=\frac{3N^2}{4\pi^2}|F_\omega|^2
        =\frac{4\pi^2}{3N^2}|\Sigma|^2\left|(3+\epsilon)\lambda\omega^{2+\epsilon}k^{-\epsilon}-F_\Sigma^*\omega^*\left|\omega/k\right|^{2\epsilon}\right|^2.
	\end{equation}
    We will soon show that $V_{F_\omega}$ is subdominant with respect to $V_{F_\Sigma}$ for the stabilization mechanism we are considering.
    
	When SUSY is unbroken in the UV $(F_\Sigma=0)$, the potential is $V_\eff\propto|\lambda|^2|\omega|^{4+2\epsilon}$ whose minimum is $\omega_{\min}=0$, and no spontaneous breaking occurs. Therefore, the SUSY-breaking coupling $F_\Sigma\neq0$ triggers the spontaneous breaking of scale invariance in our model.\footnote{If we had turned on the quartic $\kappa\neq0$, there would be a nonzero minimum that conserves SUSY. However, if $\kappa$ is sufficiently small its effects can be neglected, so we chose $\kappa=0$ for ease of discussion.} $V_{F_\Sigma}$ has a minimum only when $\epsilon>1$, and both potentials have a minimum with a parametric dependence of $|\omega_{\min}/k|^{\epsilon-1}\sim |\lambda| k / m_{3/2}$. The absolute value of the minimum should be much smaller than the UV cutoff $|\omega_{\min}|\ll k$, which is only possible if $|\lambda|\ll m_{3/2}/k$. For example, if we take $k=M_\pl$ and $m_{3/2}=10\eV$, we find that the coupling should be smaller than $|\lambda|\ll 10^{-26}$. This small value is only possible because $\lambda$ is protected by SUSY. 
 
    Near the minimum at $\omega=\omega_{\min}+\delta\omega$, the potentials are approximately \begin{equation}
        V_{F_\Sigma}\approx m_{3/2}^2\left|\frac{\omega_{\min}}{k}\right|^{2\epsilon}|\delta\omega|^2,\quad V_{F_\omega}\approx \frac{4\pi^2}{3N^2}|\Sigma|^2m_{3/2}^2\left|\frac{\omega_{\min}}{k}\right|^{4\epsilon}|\delta\omega|^2.
    \end{equation}
    $V_{F_\omega}$ is suppressed by $|\omega_{\min}/k|^{2\epsilon}$ with respect to $V_{F_\Sigma}$, so its effects are negligible. A similar analysis shows that $V_{F_\Sigma}$ dominates over $V_{F_\omega}$ for all $\omega$, except possibly near the UV cutoff $|\omega|\approx k$, where our EFT description is expected to break down regardless. We can therefore ignore $V_{F_\omega}$ in the following discussions.
 
	Now that we understand that the effective potential is $V_\eff=V_{F_\Sigma}$, we can make the previous calculation more precise. In terms of the radial $\chi$ and the angular $\vartheta$ components of the scalar modulus $\omega$ --- the dilaton and the R-axion, respectively --- the effective potential in Eq.~\eqref{eq_CFT_SUSYbrk_modulus_potential} is
	\begin{equation}\label{eq_CFT_SUSYBRK_dilaton_raxion_potential}
		V_\eff(\chi,\vartheta)=\frac{1}{1+\epsilon}m_{3/2}^2\chi^2\left(\frac{\chi}{k}\right)^{2\epsilon}-2|\lambda|m_{3/2}\chi^3\left(\frac{\chi}{k}\right)^\epsilon \cos[\arg\lambda+\arg F_\Sigma+(3+\epsilon)\vartheta].
	\end{equation}
	The minimum of this potential is at
	\begin{equation}\label{eq_CFT_dilaton_raxion_VEV}
		\chimin=\left[(3+\epsilon)|\lambda|\frac{k}{m_{3/2}}\right]^{1/(\epsilon-1)}k,\qquad \vartheta_{\rm min}=-\frac{\arg\lambda+\arg F_\Sigma}{3+\epsilon}.
	\end{equation}
    As expected, when $\lambda\ll m_{3/2}/k$ is exponentially small, we generate a large hierarchy $\chimin/k\ll1$ between the UV scale $k$ and the IR scale $\chimin$ where the CFT is spontaneously broken. It is the smallness of $\lambda$ in the superpotential (which is technically natural because it is protected by SUSY) that generates the UV/IR scale hierarchy in the SCFT.
	
	This is an unusual stabilization mechanism for the dilaton compared to previous proposals in the literature; e.g., the commonly used Goldberger-Wise mechanism~\cite{Goldberger:1999uk} relies on the CFT-breaking operator being almost marginal $(|d_\O-4|\ll1)$ with order-one coefficients. A relevant operator $(2<d_\O<4)$ can also stabilize the dilaton if the coupling of $\O$ in the UV is small due to some symmetry such as a broken $Z_2$~\cite{Csaki:2023pwy}. Here, the hierarchy is generated by an irrelevant operator $(d_\O>4)$, and this is true even if the operator is far from marginal (i.e. $\epsilon$ is not close to 1). It is the underlying SUSY that makes this stabilization mechanism possible: generically, the dilaton potential has the scale invariant $\chi^4$ term with an order-one coefficient, while irrelevant operators generate $\chi^{d_\O} k^{4-d_\O}$ terms of higher powers. When these two terms are balanced against one another, the dilaton VEV is at $\chimin\sim k$ and no large UV/IR hierarchy is generated. A supersymmetric model can avoid this argument because we can set the quartic term to zero due to the nonrenormalization theorem, as we have done here. 
	
	\subsection{Spectrum}\label{sec_spectrum}
	
	The masses of the scalar fields are computed by taking the second derivative of the scalar potential in Eq.~\eqref{eq_CFT_SUSYBRK_dilaton_raxion_potential} and plugging in the VEVs, taking into account the wavefunction normalization. We find that the mass of the dilaton is equal to
	\begin{equation}\label{eq_CFT_dilaton_mass}
		m_\chi=\sqrt{\frac{4\pi^2}{3N^2}}\sqrt{\epsilon-1}\left(\frac{\chimin}{k}\right)^\epsilon m_{3/2}=\sqrt{\frac{4\pi^2}{3N^2}}\sqrt{\epsilon-1}(3+\epsilon)|\lambda|\chimin,
	\end{equation}
	where $m_{3/2}=M_\SUSYbrk^2/M_\pl$ is the SUSY-breaking scale in the dark sector. Note that since $\lambda$ is exponentially small and $\lambda$ is protected by SUSY, the dilaton mass is much smaller than $\chimin$. This is one of the central results of our paper: our mechanism generates a light dilaton that is exponentially lighter than the IR scale. For comparison, in the Goldberger-Wise mechanism, the UV/IR hierarchy is generated by a small anomalous dimension $|d_\O-4|\ll1$, and the dilaton mass is only $|d_\O-4|^{1/2}$ smaller than $\chimin$~\cite{Csaki:2000zn,Chacko:2012sy,Bellazzini:2013fga}. We cannot lower the dilaton mass by making $|d_\O-4|$ arbitrarily small without making the UV scale $k$ trans-Planckian~\cite{Hubisz:2024hyz}. In our model the source of the hierarchies is the small parameter $\lambda$ and not the anomalous dimension $\epsilon$, which allows us to set the dilaton mass to be small without making $k>M_\pl$.

    The mass of the dilaton is warped down from the gravitino mass by $(\chimin/k)^\epsilon$ and satisfies $m_\chi\ll m_{3/2}$. Therefore even though the smallest realistic value of the gravitino mass is
    \begin{equation}
        m_{3/2}\approx \frac{M_\SUSYbrk^2}{M_\pl}\gtrsim 10\eV,
    \end{equation}
    corresponding to the smallest possible SUSY-breaking scale of $M_\SUSYbrk\approx100\TeV$ (see Sec.~\ref{sec_setup}), the dilaton can easily be lighter than $1\eV$ --- the upper limit of the ULDM mass regime. In Sec.~\ref{sec_misalignement} we analyze in detail the viability of the dilaton to be ULDM.
	
	The masses of the other fields in the modulus supermultiplet are not identical to the dilaton mass because SUSY is broken. The masses of the R-axion $\vartheta$ (the angular mode of the complex scalar $\omega$) and the dilatino $\tilde\omega$ (the fermion component) are
	\begin{equation}\label{eq_CFT_Raxion_dilatino_masses}
	\begin{aligned}
	m_\vartheta&=\sqrt{\frac{4\pi^2}{3N^2}}\sqrt{3+\epsilon}\left(\frac{\chimin}{k}\right)^\epsilon m_{3/2}\sim m_\chi,\\
	m_{\tilde\omega} &= \frac{4\pi^2}{3N^2}|\Sigma|m_{3/2}\left(\frac{\chimin}{k}\right)^{2\epsilon}\sim m_\chi \left(\frac{\chimin}{k}\right)^\epsilon.
	\end{aligned}
	\end{equation}
	Misalignment will produce both R-axions and dilatons, with each accounting for about half of the total DM abundance due to their nearly equal masses (see Sec.~\ref{sec_misalignement}). The dilatino is much lighter than its scalar partners so in principle they can decay to the dilatino, but this process is suppressed by a tiny Yukawa coupling of order $m_{3/2}/k (\chimin/k)^{2\epsilon-1}$ and is therefore very rare. The dilatino is not produced by misalignment because it is a fermion. The phenomenology of $\vartheta$ and $\tilde\omega$ is discussed in Sec.~\ref{subsec_r-axion_dilatino_pheno}.
	
\begin{figure}
		\centering
		\includegraphics[width=1\textwidth]{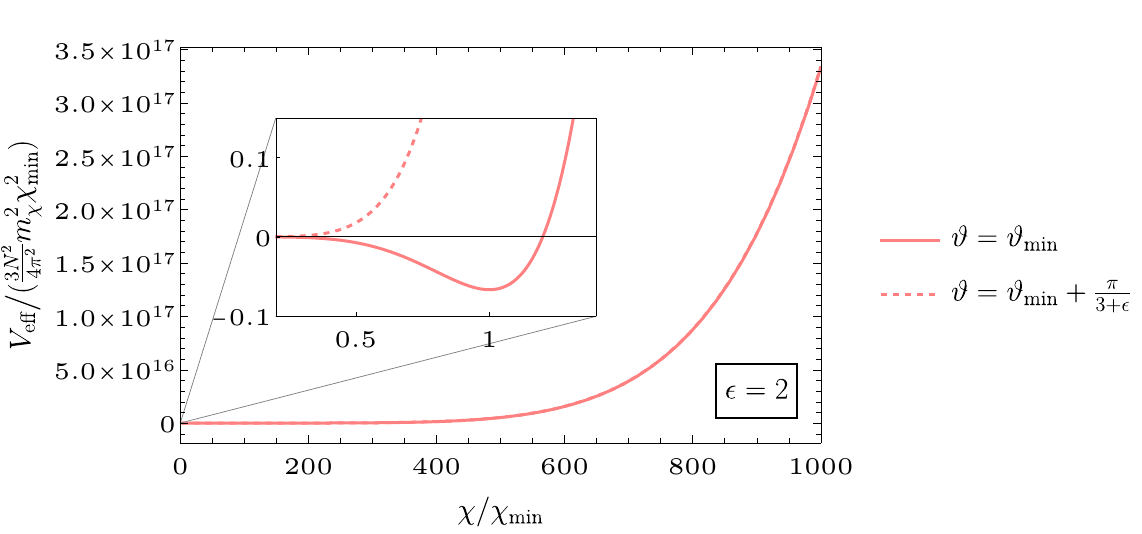}
        \includegraphics[width=0.8\textwidth]{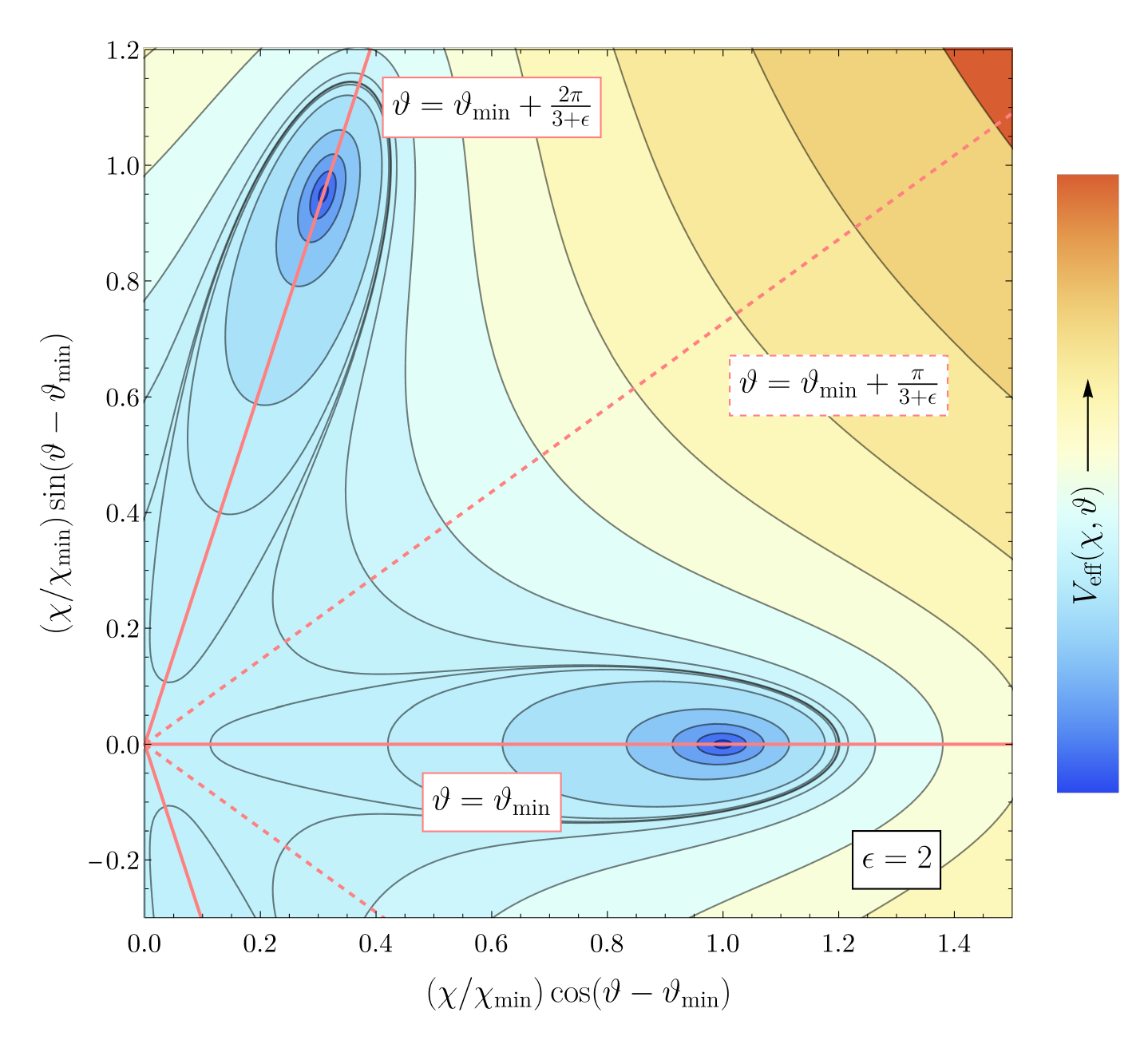}
		\caption{A plot of the modulus potential $V_\eff(\chi,\vartheta)$ as a function of the dilaton $\chi$ and the R-axion $\vartheta$, where the anomalous dimension is fixed to $\epsilon=2$. (\textit{top}) A plot of the dilaton potential $V_\eff(\chi)$ along the $\vartheta=\vartheta_{\min}$ and $\vartheta=\vartheta_{\min}+\pi/(3+\epsilon)$ directions, in solid-pink and dashed-pink respectively. The dilaton field value is normalized by its minimum value $\chimin$, and the potential is normalized as in Eq.~\eqref{eq_modulus_potential}. For large dilaton field values, the potential is essentially symmetric for all $\vartheta$ values. This symmetry is broken near the minimum as shown in the inset and in the bottom figure. (\textit{bottom}) A contour plot of the dilaton potential $V_\eff(\chi,\vartheta)$ near its minimum. The broken symmetry in $\vartheta$ is manifest.}
		\label{fig_modulus_potential}
	\end{figure}

	For later convenience, we note that the modulus potential in Eq.~\eqref{eq_CFT_SUSYBRK_dilaton_raxion_potential} can be written as
	\begin{equation}\label{eq_modulus_potential}
		V_\eff=\frac{3N^2}{4\pi^2}\frac{m_\chi^2\chimin^2}{\epsilon-1}
        \left[\frac{1}{1+\epsilon}\left(\frac{\chi}{\chimin}\right)^{2+2\epsilon}
        -\frac{2}{3+\epsilon}\left(\frac{\chi}{\chimin}\right)^{3+\epsilon}
        \cos[(3+\epsilon)(\vartheta-\vartheta_{\rm min})]\right].
	\end{equation}
	This potential is valid up to the UV cutoff $\chi\approx k$ and is plotted in Fig.~\ref{fig_modulus_potential}. The corresponding kinetic term is
	\begin{equation}\label{eq_modulus_kinetic}
		\L_{\rm kin}=\frac{3N^2}{4\pi^2}|\partial_\mu\omega|^2
        =\frac{3N^2}{4\pi^2}\left[(\partial_\mu\chi)^2+\chi^2(\partial_\mu\vartheta)^2\right].
	\end{equation}
    The canonically normalized dilaton is $\overline\chi=\sqrt{3N^2/2\pi^2}\chi$ and the corresponding VEV is $f\equiv\sqrt{3N^2/2\pi^2}\chimin$, which is the SSB scale of the CFT.

	\subsection{Naturalness Bound from AMSB}\label{sec_AMSB}
	
	There is a lower limit on how small we can make $m_\chi$ before radiative corrections become comparable to the bare mass in Eq.~\eqref{eq_CFT_dilaton_mass} and the model becomes unnatural. This is because although $\lambda$ is protected from radiative corrections by SUSY, higher-dimensional operators induced by SUSY-breaking will generate corrections to $m_\chi$ that are potentially larger. The two dominant sources of SUSY-breaking in our model are the running of $F_\Sigma$ from the UV scale --- which is used to stabilize and give the dilaton its mass --- and through anomaly mediation (AMSB)~\cite{Randall:1998uk,Giudice:1998xp}. The effects of the latter dictate the naturalness bound on our model.
	
	AMSB is dominant over gravity-mediated SUSY-breaking because the dark sector is sequestered from the hidden sector wherein SUSY is broken. This is a critical assumption in our model, since Planck-suppressed contact terms between the dark sector and the hidden sector would generically induce $O(m_{3/2})$ corrections to the dilaton mass and spoil its smallness. Sequestering suppresses these contact terms. This suppression can be explained in either the CFT~\cite{Luty:2001jh,Luty:2001zv,Schmaltz:2006qs} or the 5D pictures~\cite{Randall:1998uk}; in the CFT picture, the Planck-suppressed contact terms are a small perturbation to the UV couplings. The CFT flows to an IR fixed point that is insensitive to the UV initial conditions, so these contact terms obtain a large anomalous dimension due to the CFT dynamics and their values in the IR are consequently suppressed. In the 5D picture, the hidden sector and the DS are localized on the UV and the IR branes, respectively, so contact terms are suppressed by a small wavefunction overlap between the fields on the two branes.
	
	Even in the absence of direct couplings between the hidden sector and the DS induced by graviton loops, SUSY-breaking is still transmitted by the scalar auxiliary field $F_\varphi$ of the off-shell supergravity (SUGRA) multiplet. It obtains a VEV $F_\varphi\approx m_{3/2}$ due to SUSY-breaking and its effects at low energies are captured by the conformal compensator $\bs\varphi=1+\theta^2F_\varphi$~\cite{Randall:1998uk,Giudice:1998xp}. The conformal compensator is used in SUGRA as a spurion that restores scale invariance to the theory, and whose lowest component breaks superconformal symmetry down to super-Poincar\'e symmetry. By dimensional analysis $\bs\varphi$ couples to the K\"ahler potential as $\int\odif[order=4]{\theta}\bs\varphi^\dagger\bs\varphi K$, and to the superpotential as $\int\odif[order=2]{\theta}\bs\varphi^3W$. This is called anomaly mediation because $F_\varphi$ couples to all mass/anomalous dimensions in the Lagrangian.

    When we restore the conformal compensator $\bs\varphi$ to the Lagrangian in Eq.~\eqref{eq_CFT_low_energy_SUSY_modulus_EFT},
    \begin{equation}
        \L_\eff=\int\odif[order=4]{\theta}\bs\varphi^\dagger\bs\varphi\bs\omega^\dagger\bs\omega
        \left[\frac{3N^2}{4\pi^2}-\frac{1}{1+\epsilon}\left|\bs\Sigma\left(\frac{\bs\omega}{k}\right)^\epsilon\right|^2\right]
        +\int\odif[order=2]{\theta}\lambda\bs\varphi^3\bs\omega^3\bs\Sigma\left(\frac{\bs\omega}{k}\right)^\epsilon+\hc
	\end{equation}
    we see that its effects can be absorbed into $F_\omega$ and $F_\Sigma$ by the redefinitions $\bs\varphi\bs\omega\rightarrow\bs\omega$ and $\bs\Sigma\bs\varphi^{-\epsilon}\rightarrow\bs\Sigma$. This degeneracy is expected because $\bs\Sigma$ is also a spurion that restores the scale invariance of the SCFT that was broken by the operator $\bs\O$. Since $F_\varphi\approx F_\Sigma\approx m_{3/2}$, we see that AMSB has no effect on the preceding discussions. However, in addition to the K\"ahler corrections in Eq.~\eqref{eq_CFT_low_energy_SUSY_modulus_EFT} arising from the CFT dynamics, we expect higher-dimensional interactions of the dilaton with the graviton to generate additional K\"ahler corrections of the form
	\begin{equation}
		\L_\eff\supset\int\odif[order=4]{\theta}\bs\varphi^\dagger\bs\varphi
        \left[\bs\omega^\dagger\bs\omega-\frac{1}{16\pi^2}\frac{(\bs\omega^\dagger\bs\omega)^2}{M_\pl^2}\right]
        \xrightarrow{\bs\varphi\bs\omega\rightarrow\bs\omega}
        \int\odif[order=4]{\theta}\left[\bs\omega^\dagger\bs\omega-\frac{1}{16\pi^2}\frac{(\bs\omega^\dagger\bs\omega)^2}{\bs\varphi^\dagger\bs\varphi M_\pl^2}\right],
	\end{equation}
    where we have canonically normalized $\bs\omega$ in this expression for simplicity.	In the SCFT picture this correction is expected because conformal invariance is broken by gravity, and in the dual 5D picture this correction arises because the graviton propagates in the bulk~\cite{Luty:2002ff}. Its contribution to the modulus scalar mass is roughly
	\begin{equation}\label{eq_naturalness_bound_1}
		\Delta m_\omega^2 \sim \frac{1}{16\pi^2}\frac{f^2}{M_\pl^2}|F_\varphi|^2\sim \frac{1}{16\pi^2}\frac{f^2}{M_\pl^2}m_{3/2}^2.
	\end{equation}
	The validity of our analysis thus requires $m_\chi^2,m_\vartheta^2\gtrsim\Delta m_\omega^2$, which is the naturalness bound of our model. As we will see, this bound excludes a significant fraction of our parameter space, however leaving ample allowed space.
	
	\section{Misalignment Production}\label{sec_misalignement}

    We now have a consistent model of an ultralight dilaton, and can therefore explore its possible production mechanisms as a DM candidate. Here we focus on the misalignment mechanism~\cite{Dine:1982ah,Preskill:1982cy,Abbott:1982af} for the dilaton and the R-axion; essentially, the particles are in a coherent superposition and act as a classical field whose energy is determined by the initial displacement from the potential minimum. In the early Universe, the field value remains constant due to Hubble friction. As the Universe expands and Hubble decreases, the field begins to roll down the potential until it eventually oscillates about the minimum. The energy density of the oscillating field redshifts like matter, consistent with cosmological observations of a CDM component in the Universe~\cite{Kuhlen:2012ft}.
    
    The standard ULDM candidate analyzed in the misalignment mechanism is the axion/ALP~\cite{Hook:2018dlk}, whose potential is periodic and can be approximated as harmonic (i.e. quadratic) for most of the field range. The dilaton, however, is not a periodic field, and its potential is very anharmonic for most of the field range (see Eq.~\eqref{eq_modulus_potential} and Fig.~\ref{fig_modulus_potential}). As a result, for most possible initial conditions anharmonic effects substantially alter the final dilaton abundance. We will be agnostic about the initial field value $\chi_0$ and study the misalignment production in two limits of interest: the $\chi_0\ll\chimin$ regime near the origin in Sec.~\ref{sec_misalignment_initial_origin} and the $\chi_0\gg\chimin$ regime far away from the minimum in Sec.~\ref{sec_misalignment_initial_k}.

    We additionally assume that the dark sector is not reheated at the end of inflation. This assumption is crucial for two reasons. The first is that ULDM was necessarily never in thermal equilibrium, since it is below the warm DM limit of $1\keV$. The second reason is that if the CFT were sufficiently heated to be in the deconfined phase, then due to the shallow dilaton potential, the phase transition to the confined phase would be extremely suppressed and never complete~\cite{Creminelli:2001th}. Then the Universe would be stuck in the deconfined phase with a positive CC and inflate eternally.
    
    To analyze the production of the ULDM dilaton through misalignment, it is convenient to rewrite the effective action in terms of a dimensionless dilaton field,
	\begin{equation}
		\phi = \frac{\chi}{\chi_{\rm min}},
	\end{equation}
	and to shift the axion $\vartheta \rightarrow \vartheta + \vartheta_{\rm min}$.
	Eqs.~\eqref{eq_modulus_potential} and~\eqref{eq_modulus_kinetic} then give the effective Lagrangian
	\begin{equation}\label{eq:lagrangian}
		\L_\eff= \frac{1}{2}f^2 \left[(\partial_\mu\phi)^2+\phi^2(\partial_\mu\vartheta)^2 - \frac{m_\chi^2}{\epsilon-1}\left(\frac{\phi^{2+2\epsilon}}{1+\epsilon}-\frac{2\phi^{3+\epsilon}}{3+\epsilon}\cos[(3+\epsilon)\vartheta]\right)\right]+V_{\min}.
	\end{equation}
    where we have shifted the potential by $V_{\min}\equiv-\frac{1}{2}m_\chi^2f^2/(\epsilon^2+4\epsilon+3)$ in order to set the CC to zero at the minimum.
	The axion decay constant is $f/ (3+\epsilon)$, which is reflected in the periodicity of the potential under $\vartheta \rightarrow \vartheta + 2\pi/(3+\epsilon)$. This potential admits a unique minimum at $\phi = 1$, $\vartheta = 0$. To study the misalignment mechanism, we consider some initial field values $\phi_0, \vartheta_0$ (with $\phi_0 > 0$, $0 \leq \vartheta_0 < 2\pi/(3+\epsilon)$, and zero initial velocity) and compute their time evolution through the equations of motion on an FRW metric.

    Although we will study all possible initial conditions in the next sections, there is a minimal prediction for the dilaton initial condition where no additional assumptions are made about the dilaton dynamics before the radiation domination (RD) era. To see this, observe that because the dilaton has a universal coupling to the trace of the stress-energy tensor (cf. Eq.~\eqref{eq:dilaton_couplings_general} in Sec.~\ref{sec_dilaton_phenomenology}), it couples to the total energy density of the Universe and thus to the Hubble parameter $H$ of the Universe. Using the Friedmann equations and assuming that $w=P/\rho$ is the equation-of-state parameter of the Universe, this coupling induces an effective mass for the dilaton of the form
    \begin{equation}\label{eq_Hubble_mass_dilaton_potential}
        V_H=\frac{\braket{T^\mu_{\rm{m}\mu}}}{12M_\pl^2}\overline\chi^2=\frac{1-3w}{4}H^2\overline\chi^2,
    \end{equation}
    where $\overline\chi$ is the canonically normalized dilaton field. We refer to this as the ``Hubble mass'' contribution to the dilaton potential. In the RD era the Hubble mass nearly vanishes because $w=1/3$ to leading order,\footnote{We have verified that small deviations from $w=1/3$ during RD are insignificant for the dilaton dynamics in the parameter space of interest.} and in the later eras the Hubble mass is small and negligible compared to the DM mass, $H\lesssim H_{\rm eq}\approx10^{-28}\eV\ll m_{\rm ULDM}$. The Hubble mass can be significant before RD when $w\neq1/3$ and Hubble is large --- especially during inflation.
    
    Indeed, if there are no dilaton-inflaton couplings beyond those in Eq.~\eqref{eq_Hubble_mass_dilaton_potential} above, then during inflation the Hubble mass is the dominant contribution to the dilaton potential\footnote{This is certainly true when comparing the Hubble mass with $V_\eff$ in Eq.~\eqref{eq_CFT_SUSYBRK_dilaton_raxion_potential} when $H_I\gtrsim m_{3/2}$. For smaller values of $H_I$, the potential in Eq.~\eqref{eq_CFT_SUSYBRK_dilaton_raxion_potential} is steeper and thus accelerates the rolling of the dilaton to the origin relative to the rolling induced by the Hubble mass alone. The Hubble mass becomes  dominant  below $\chi\sim k (H_I/m_{3/2})^{1/\epsilon}$.} with $m_{\chi,\eff}^2=2H_I^2$ where $H_I$ denotes the inflationary Hubble scale. After solving the EOM we find that during inflation the dilaton rolls exponentially with the number of e-folds $N_e\equiv H_It$ as $\overline\chi\sim e^{-N_e}$, implying that $\overline\chi\lesssim10^{-22}M_\pl$ by the end of inflation since at least 50 e-folds of inflation are required from CMB observations~\cite{Planck:2018jri}. Therefore, the initial $\overline\chi_0$ at the onset of RD is $\overline\chi_0\lesssim10^5\eV$ for the minimal scenario. However, nonminimal couplings between the inflaton and the dilaton could change this prediction and allow for a broader range of $\overline\chi_0$ values, so we have chosen to calculate the dilaton production for all initial field values.
	
    \subsection{Initial condition $\chi_0 \ll \chi_{\rm min}$}\label{sec_misalignment_initial_origin}

    We first compute the relic abundance when the initial field value $\phi_0=\chi_0 /\chi_{\rm min} \ll 1$ is close to the origin. In this case, one has almost standard misalignment: an initial overdamped phase where the energy density is constant, followed by underdamped harmonic oscillations about the potential minimum where the energy density redshifts like cold matter. The only difference from the standard scenario is that the time $t_\osc$ at which the dilaton begins to oscillate depends on its initial value. This is due to the anharmonic potential near the origin, $V\sim\phi^{3+\epsilon}$ [cf. Eq.~\eqref{eq:lagrangian}]. To see this in more detail, note that the equation of motion for $\phi$ near the origin is
    \begin{equation}
        \ddot{\phi} + 3 H \dot{\phi} - \frac{m_\chi^2}{\epsilon - 1} \phi^{2+\epsilon}\cos[(3+\epsilon)\vartheta]  = 0 ,
    \end{equation}
    where we neglect the motion of the R-axion field for simplicity. The dilaton field begins to roll when the potential gradient term dominates over the Hubble friction term ($H=1/2t$ during RD), leading to
    \begin{equation}\label{eq_tcrit}
        t_\osc \sim \frac{3}{2} \sqrt{ \frac{\epsilon - 1}{\phi_0^{1+\epsilon}} } \times \frac{1}{m_\chi}\,.
    \end{equation}
    Thus the onset of the oscillations is delayed compared to the case of a purely quadratic potential where the oscillations would start at $t_\osc\sim 1/m_\chi$. 
    This is similar to the case of anharmonic effects for the QCD axion hilltop misalignment~\cite{Turner:1985si,Lyth:1991ub,Strobl:1994wk,Bae:2008ue,Visinelli:2009zm}.
    
    The initial energy density is $-V_{\min}$, corresponding to the potential at small $\phi$. Explicitly, the DM energy density is then given by
	\begin{equation}
		\frac{\rho(t)}{\frac{1}{2}m_\chi^2 f^2} =
		\frac{1}{\epsilon^2+4\epsilon+3}\times\begin{cases}
			1 & t < t_\osc \quad {\rm (overdamped)} \\
			(t/t_\osc)^{-3/2} & t_\osc \leq t   \quad {\rm (harmonic)}
		\end{cases}
	\end{equation}
    For a generic initial condition, this energy would be divided equally between the dilaton and the R-axion due to their nearly equal masses [cf. Eq.~\eqref{eq_CFT_Raxion_dilatino_masses}]. Using entropy conservation and the fact that the oscillations begin during the RD era, we can calculate the final dilaton DM relic abundance today, $\Omega_\DM = \rho/\rho_c$, as follows~\cite{Chatrchyan:2023cmz}:
    \begin{align}
        \frac{\Omega_{\rm DM} h^2}{0.1} &
        \sim \frac{1}{0.1}\times \frac{1}{3} \frac{g_{\star s}(T_0)}{g_{\star s}(T_\osc)}
        \left[\Omega_\gamma \frac{g_\star(T_\osc)}{g_\star(T_0)}\right]^{4/3}\frac{\rho(t_\osc)}{M_\pl^2H_0^{1/2}H_\osc^{3/2}}\label{eq_misalignment_general_DM_abundance}\\
        &\sim \frac{c_1(\epsilon)}{c_1(2)}  \left(\frac{\phi_0}{10^{-5}}\right)^{-3(1+\epsilon)/4} \times \left( \frac{m_\chi}{1{\rm~eV}} \right)^{1/2} \left( \frac{ f }{2 \times 10^{6}\GeV} \right)^2,\label{eq_misalignment_origin_DM_abundance}
    \end{align}
    where the prefactor is $c_1(\epsilon)=(\epsilon-1)^{3/4}/(\epsilon^2+4\epsilon+3)$. This differs by a factor of $1/(\epsilon^2 + 4\epsilon + 3) \times (m_\chi t_\osc)^{3/2}$ from the usual expression for misalignment production of DM with an $O(1)$ misalignment angle~\cite{Marsh:2015xka}, where the dominant difference is due to the late oscillation time $t_\osc \gg 1/m_\chi$.
    
    The matter power spectrum is known to follow the standard $\Lambda$CDM cosmology up to $k$ modes of order $10~\rm{Mpc}^{-1}$~\cite{Kuhlen:2012ft}, corresponding to modes that entered the horizon after $T_\gamma\approx \keV$. This constrains the oscillation time $t_\osc$, at which the dilaton and the R-axion begin to redshift like matter, to occur before the Universe reaches this temperature: $T_\gamma(t=t_\osc)\gtrsim \keV$. This gives the bound
    \begin{equation}\label{eq_structure_formation_bound}
        H_\osc \sim \frac{2}{3} \sqrt{ \frac{\phi_0^{1+\epsilon}}{\epsilon - 1} }\times m_\chi \gtrsim \frac{\keV^2}{M_\pl}\simeq 4.2 \times 10^{-22}\eV,
    \end{equation}
    where Eq.~\eqref{eq_tcrit} was used.

    The dilaton field experiences de Sitter fluctuations during inflation that are typically of magnitude $\delta \phi = H_I/2\pi f$, which are independent of the inflaton fluctuations. Since in the present scenario $\phi_0\ll1$, isocurvature fluctuations are produced with magnitude~\cite{Kobayashi:2013nva,Schmitz:2018nhb}
    \begin{equation}
        \left|\frac{d \log \Omega_{\rm DM}}{d \phi_0} \delta \phi\right| = \frac{3(1+\epsilon) H_I}{8\pi \overline{\chi}_0} \lesssim 3 \times 10^{-5}.
    \end{equation}
    The bound on this quantity is taken from Planck~\cite{Planck:2018jri}. The minimal scenario, where $\overline\chi_0\lesssim10^5\eV$, is therefore only compatible with low-scale inflation models with $H_I\lesssim O(10)\eV$. This is not true in the general case where larger $\overline\chi_0$ values are possible.
    
    \subsection{Initial condition $\chi_0 \gg \chimin$}\label{sec_misalignment_initial_k}
    
	We now compute the DM relic abundance produced by misalignment when the initial field value $\phi_0 =\chi_0/\chi_{\rm min} \gg 1$ is far from the minimum of the potential. In this case, the evolution of the energy density over time is separated into three distinct phases:
	\begin{itemize}
		\item \textbf{Overdamped phase:} at early times Hubble friction freezes the fields at their initial values $\phi = \phi_0$, $\vartheta = \vartheta_0$. The energy density is constant, corresponding to an equation-of-state parameter $w = -1$.
		\item \textbf{Anharmonic phase:} at a time $t_\anh \sim \phi_0^{-\epsilon}/m_\chi$, the fields roll away from their initial values and begin large, anharmonic oscillations. The energy density redshifts with $w = \epsilon/(2+\epsilon)$.
		\item \textbf{Harmonic phase:} at a time $t_\osc \sim \phi_0^{(4-\epsilon)/3}/m_\chi$, the fields have dissipated enough energy to begin underdamped harmonic oscillations about the potential minimum at $\phi = 1, \vartheta = 0$. The energy density now behaves as CDM ($w = 0$).
	\end{itemize}
	The overdamped period followed by underdamped harmonic oscillations is the typical ALP misalignment story, but the intermediate phase of anharmonic oscillations is perhaps less familiar. It arises because the dilaton field is not compact and does not have a periodic potential, so it is not valid to approximate it as quadratic for large field values $\phi \gg 1$. During the anharmonic phase, the DM undergoes oscillations in a $\phi^{2+2\epsilon}$ potential, causing the energy density to dilute faster than radiation with $w = \epsilon/(2+\epsilon)$ (which was first shown in Ref.~\cite{Turner:1983he}). The anharmonic phase ends when the energy density becomes small enough that the oscillations are confined to the vicinity of the minimum at $\phi = 1, \vartheta = 0$, where the potential is approximately quadratic and the oscillations are therefore harmonic.

    This description accurately captures the behavior for $\epsilon < 4$; for $\epsilon \geq 4$, the behavior qualitatively changes. Due to the steeper potential, the dilaton rolls down fast enough to ``catch up'' with the decreasing Hubble friction, which temporarily slows down the field until Hubble becomes sufficiently small and the cycle repeats. Instead of oscillating, the dilaton rolls down in steps until it reaches the vicinity of the minimum and is frozen there by Hubble friction. There, the anharmonic shape of the potential delays the onset of oscillations, and the behavior is qualitatively similar to the scenario considered in Sec.~\ref{sec_misalignment_initial_origin}, albeit initially closer to the minimum rather than localized near the origin. Therefore, we limit our study of the parameter space for this scenario to $\epsilon < 4$.
	
\begin{figure}
		\centering
		\includegraphics[width=0.8\textwidth]{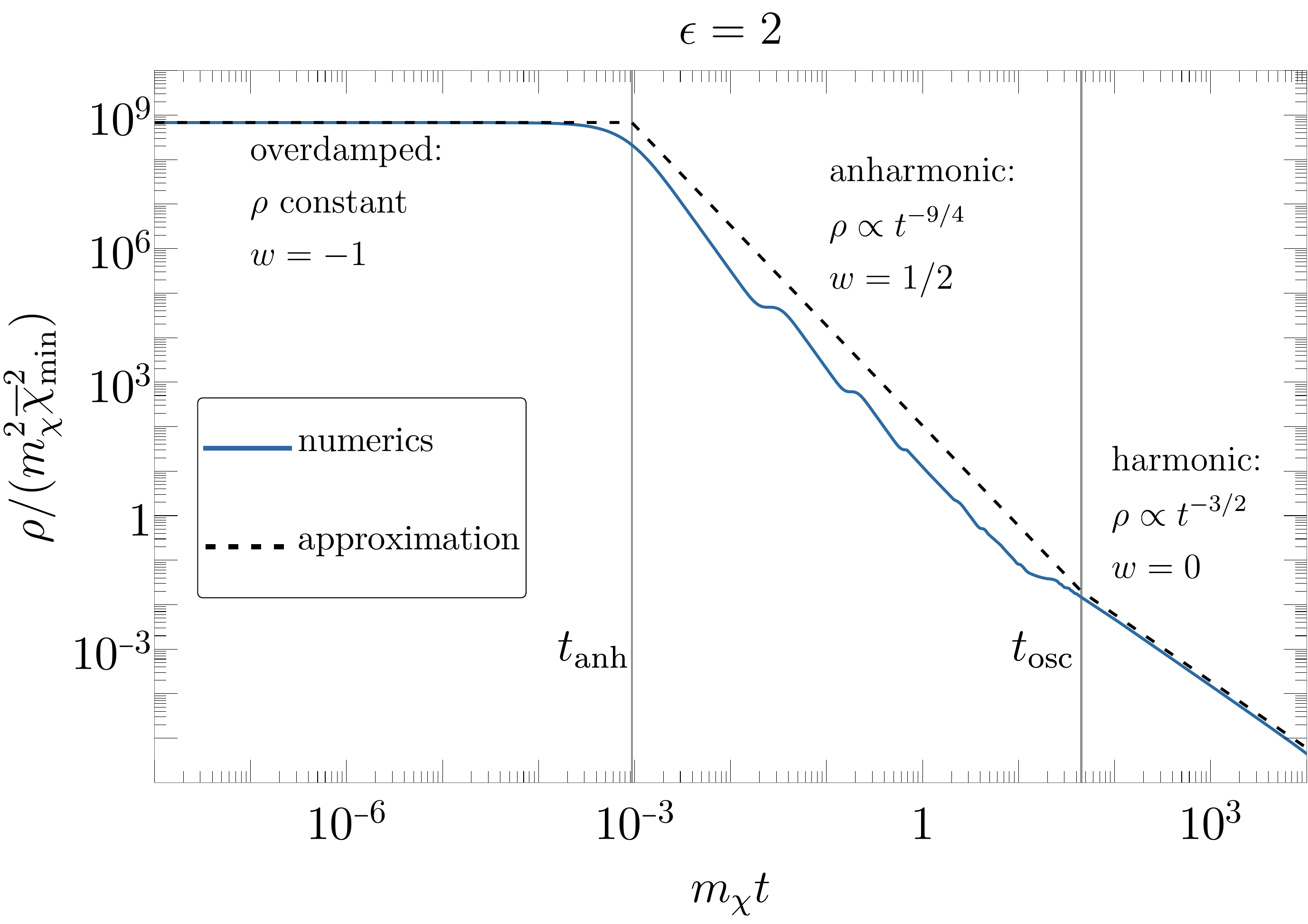}
		\caption{Time evolution of the DM field energy density $\rho$ with the initial condition $(\phi_0, \vartheta_0) = (40, 1)$ and $\epsilon = 2$. We show a numerical result obtained by solving the equations of motion in a radiation-dominated FRW universe~\textit{(solid blue line)} and the analytical approximation described in Eq.~\eqref{eq:energydensity}~\textit{(dashed black line)}. The behavior separates into the three distinct regimes described in the main text; for each we provide the equation-of-state parameter $w$ and the time-dependence of the energy density. We indicate the timescales $t_\anh$ and $t_\osc$, computed with Eq.~\eqref{eq:timescales}, which correspond to transitions between the different regimes.}
		\label{fig:misalignment}
	\end{figure}

	The derivation of the timescales $t_\anh,t_\osc$ and the scaling of the energy density during the anharmonic oscillations is straightforward, and we include it in App.~\ref{app_energy_density}. The upshot is that the energy density during radiation domination is approximated by
	\begin{equation}\label{eq:energydensity}
		\frac{\rho(t)}{\frac{1}{2}m_\chi^2f^2} =
		\begin{cases}
			\dfrac{\phi_0^{2+2\epsilon}}{\epsilon^2 - 1} & t < t_\anh \quad {\rm (overdamped)} \\
			\dfrac{\phi_0^{2+2\epsilon}}{\epsilon^2 - 1} \times (t/t_\anh)^{-3(1+\epsilon)/(2+\epsilon)} & t_\anh \leq t < t_\osc \quad {\rm (anharmonic)} \\
			\dfrac{1}{(3+\epsilon)^2} \times (t/t_\osc)^{-3/2} & t_\osc \leq t   \quad {\rm (harmonic)}
		\end{cases}
	\end{equation}
	where
	\begin{equation}\label{eq:timescales}
		t_\anh = \frac{3(\epsilon-1)}{2} \phi_0^{-\epsilon} \times \frac{1}{m_\chi}, \quad t_\osc = \left[ \frac{(3+\epsilon)^2}{(\epsilon^2-1)}\phi_0^{2+2\epsilon} \right]^{(2+\epsilon)/(3+3\epsilon)} t_\anh .
	\end{equation}
	Normalized by $\frac{1}{2}m_\chi^2f^2$, the initial energy density is $\phi_0^{2+2\epsilon}/(\epsilon^2 - 1)$ and $1/(3+\epsilon)^2$ is the dilaton energy at the point where the potential is well approximated by a quadratic (i.e., when the cubic term in the Taylor series expansion of the potential becomes smaller than the quadratic term). In Fig.~\ref{fig:misalignment} we compare this approximation with a direct calculation of $\rho(t)$ from the numerical solution of the EOM and find good agreement.
 
    Computing the DM relic abundance using Eq.~\eqref{eq_misalignment_general_DM_abundance}, we find
	\begin{equation}\label{eq:relicabundance1}
		\frac{\Omega_{\rm DM} h^2 }{0.1} \sim \frac{c_2(\epsilon)}{c_2(2)} \left(\frac{\phi_0}{10^5}\right)^{2-\epsilon/2} \times \left( \frac{m_\chi}{1{\rm~eV}} \right)^{1/2} \left( \frac{f}{2 \times 10^9\GeV} \right)^2 ,
	\end{equation}
	where $c_2(\epsilon)$ is
    given by
	\begin{equation}
		c_2(\epsilon) = \left( \epsilon -1 \right)^{3/2} \left( \epsilon^2 - 1 \right)^{-(2+\epsilon)/(2+2\epsilon)} \left(\epsilon + 3 \right)^{-\epsilon/(1+\epsilon)} .
	\end{equation}
    As before in Sec.~\ref{sec_misalignment_initial_origin}, the expression for the final abundance of DM is multiplied by a factor of $(m_\chi t_\osc)^{3/2}/(3+\epsilon)^2$ with respect to the abundance produced by a standard ALP misalignment, and the oscillations start much later at $t_\osc\sim \phi_0^{(4-\epsilon)/3}/m_\chi\gg 1/m_\chi$.
    
	It is important to verify that the energy density of the Universe is never dominated by the DM fields before they begin redshifting like matter, otherwise we modify the cosmological evolution of the Universe. This is guaranteed in the overdamped phase if the initial dilaton energy density is subdominant to the radiation energy density $3 H_\anh^2 M_\pl^2$ at the onset of the anharmonic oscillations, leading to a constraint on the initial condition
	\begin{equation}
		\frac{3 (\epsilon-1)}{2 (\epsilon+1)} \left( \frac{\overline\chi_0}{M_\pl} \right)^2 \lesssim 1.
	\end{equation}
    This is trivially satisfied because $\overline\chi_0\lesssim \overline k\lesssim M_\pl$ where $\overline k\equiv\sqrt{3N^2/2\pi^2}k$ . Later in the anharmonic phase, the dilaton energy redshifts faster than radiation and is therefore always subdominant. 

    As in the previous scenario, the matter power spectrum constrains the oscillation time $t_\osc$ when the dilaton and R-axion start to behave like cold matter. In analogy to Eq.~\eqref{eq_structure_formation_bound}, we have the bound
    \begin{equation}\label{eq_structure_formation_bound2}
        H_\osc=\frac{1}{3(\epsilon-1)}\left[\frac{\epsilon^2-1}{(\epsilon+3)^2}\right]^{\frac{2+\epsilon}{3+3\epsilon}} \phi_0^{(\epsilon-4)/3}m_\chi\gtrsim 4.2 \times 10^{-22}\eV
    \end{equation}
    where Eqs.~\eqref{eq:timescales} and~\eqref{eq_CFT_dilaton_mass} were used.
	
	\section{Phenomenology}\label{sec_phenomenology}
	
	\subsection{The dilaton}\label{sec_dilaton_phenomenology}

    In this section we explore the phenomenological signatures of the ultralight dilaton DM and its experimental constraints. These are determined by the coupling of the dilaton to the SM, which we calculate here. In our setup the SM sector is sequestered from the dark sector and the SM fields are elementary, rather than composite as in the original RS model~\cite{Randall:1999ee}. Consequently, the dilaton --- a composite state of the CFT --- couples very weakly to the SM, with a suppression scale $\Lambda$ much larger than the SSB scale $f$. 

    We can explicitly compute the dilaton-SM couplings in the 5D picture. The SM is UV-localized and has a small wavefunction overlap with the IR-localized dilaton, so the couplings are suppressed. To calculate them, we analyze the effect of the dilaton on the 4D Planck scale after integrating out the extra dimension, resulting in the Jordan frame action
	\begin{equation}\label{eq:eff4Dgravityaction}
		S_\eff\supset\frac{M_5^3}{2k}\int \odif[order=4]{x} \sqrt{-g}  \left[\left(\frac{\chi}{k}\right)^2-1\right] \R - \frac{M_0^2}{2}\int \odif[order=4]{x}\sqrt{-g} \R ,
	\end{equation}
	where the second term is the contribution of a UV brane-localized Einstein-Hilbert term. We identify the 4D Planck scale as $M_{\rm Pl}^2 = M_5^3/k + M_0^2$ (in our convention $M_{\rm Pl} = 1/\sqrt{8 \pi G} \approx 2.4 \times 10^{18}$~GeV).
    We can go to the Einstein frame with the usual Einstein-Hilbert action by performing a Weyl transformation $g_{\mu\nu} \rightarrow \Omega^2 g_{\mu\nu}$ with the scaling function
	\begin{equation}\label{eq_Weyl_transformation}
		\Omega^2 = M_{\rm Pl}^2 \left( \frac{M_5^3}{k}\left[1 - \left(\frac{\chi}{k}\right)^2\right] +M_0^2 \right)^{-1}  = 1 + \frac{M_5^3}{k M_{\rm Pl}^2} \left(\frac{\chi}{k}\right)^2 + O(\chi^4) .
	\end{equation}
    This Weyl transformation induces a coupling of the dilaton to the stress-energy tensor of the matter fields through its variation of the matter action $S^{\rm m}$:
	\begin{equation}\label{eq:dilaton_couplings_general}
		\delta S^{\rm m} \rightarrow \delta g_{\mu\nu} \fdv{S^{\rm m}}{g_{\mu\nu}} = -\frac{M_5^3}{k M_{\rm Pl}^2} \left(\frac{\chi}{k}\right)^2 g_{\mu\nu} \cdot \frac{\sqrt{-g}}{2} T^{\mu\nu}_{\rm m} = -\sqrt{-g} \frac{\overline{\chi}^2}{12 M_{\rm Pl}^2} T^\mu_{{\rm m}\mu}.
	\end{equation}
	In the last equality we have written the interaction in terms of the canonically normalized dilaton field $\overline{\chi} = \sqrt{6 M_5^3 /k^3} \chi$.
	Expanding about the dilaton VEV, we see that the dilaton couples linearly to the trace of the stress-energy tensor suppressed by the trans-Planckian scale $\Lambda$, i.e.
    \begin{equation}\label{eq_pheno_dilaton_SM_trace}
        \L_{\chi-\UV}=\frac{\overline{\chi}}{\Lambda} T^\mu_\mu,\quad\textrm{where }\Lambda=\frac{6 M_{\rm Pl}^2}{f}.
    \end{equation}
    Assuming that the sign of the brane-localized Ricci term in Eq.~\eqref{eq:eff4Dgravityaction} is positive, $M_0^2 > 0$, leads to an upper bound $\overline{k} < \sqrt{6} M_{\rm Pl}$. Consequently, the coupling is bounded as $\Lambda > \sqrt{6} M_{\rm Pl}$. The same suppression scale can also be found in the CFT picture~\cite{Hubisz:2024hyz}.

    The most important couplings derived from Eq.~\eqref{eq_pheno_dilaton_SM_trace} for experiments are the electron and photon couplings:\footnote{It is traditional to normalize the dilaton interactions as in~\cite{Damour:2010rp}:
	\begin{equation}
		\frac{\overline{\chi}}{\sqrt{2} M_{\rm Pl}} \left( \frac{1}{4} d_e F_{\mu\nu}F^{\mu\nu} + d_{m_e} m_e \overline{e} e\right).
	\end{equation}
	In this normalization, our dilaton couplings are given by
	\begin{equation}
		d_{m_e} = \sqrt{2} \frac{M_{\rm Pl}}{\Lambda}, \quad d_e = \frac{2 \beta(e)}{e} d_{m_e} .
	\end{equation}
    }
 	\begin{equation}\label{Eq:chi_SM}
		  \L_{\chi-\rm{SM}} \supset \frac{\overline{\chi}}{\Lambda}
          \left(m_e \overline{e} e + \frac{\beta(e)}{2e} F_{\mu\nu}F^{\mu\nu} \right),
	\end{equation}
	where $m_e$ is the electron mass and $\beta(e)$ is the QED beta function. The action for a gauge field is classically scale-invariant, so the trace of the photon stress-energy tensor is proportional to the trace anomaly.
    
    We fix the IR scale $f$ by requiring the abundance of the produced ultralight dilatons to match the observed abundance of DM via Eq.~\eqref{eq_misalignment_origin_DM_abundance} or~\eqref{eq:relicabundance1}, corresponding to the two choices of initial conditions for the dilaton misalignment analyzed in Sec.~\ref{sec_misalignement}. The suppression scale of the dilaton couplings to the SM is thus
	\begin{equation}
		\Lambda \sim 6 M_\pl\times\begin{cases}
		     \displaystyle10^{12}\sqrt\frac{c_1(\epsilon)}{c_1(2)} \left( \frac{m_\chi}{1{\rm~eV}} \right)^{1/4} \left(\frac{\chi_0/\chi_{\min}}{10^{-5}}\right)^{-3(1+\epsilon)/8}& ,\chi_0\ll\chi_{\min}\\
            \displaystyle10^9\sqrt\frac{c_2(\epsilon)}{c_2(2)} \left( \frac{m_\chi}{1{\rm~eV}} \right)^{1/4} \left(\frac{\chi_0/\chi_{\min}}{10^5}\right)^{1-\epsilon/4}& ,\chi_0\gg \chi_{\min}
		\end{cases}.
	\end{equation}
    We take the ratio $\chi_0/\chi_{\min}$ as a free parameter.
	
	\begin{figure}
		\centering
		\includegraphics[width=0.8\textwidth]{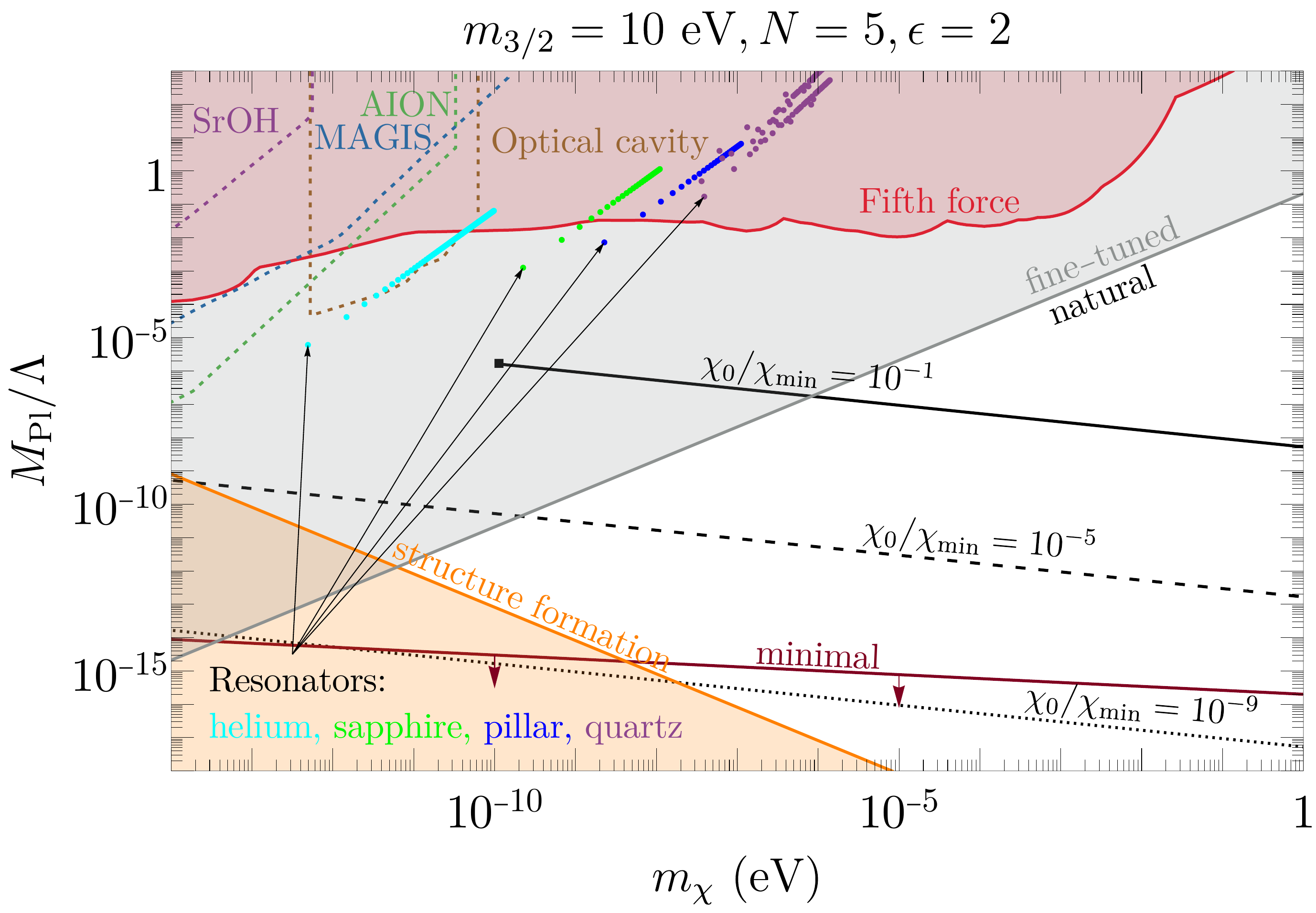}
		\caption{Exclusion plot for our ultralight dilaton with initial condition $\chi_0 \ll \chimin$, corresponding to the scenario in Sec.~\ref{sec_misalignment_initial_origin}. The black lines show the dilaton parameter space for different choices of the initial condition: $\chi_0/\chimin = 10^{-1}$~\textit{(solid black line)}, $\chi_0/\chimin = 10^{-5}$~\textit{(dashed black line)}, and $\chi_0/\chimin = 10^{-9}$~\textit{(dotted black line)}. We fix the gravitino mass $m_{3/2} = 10$~eV, the number of colors $N = 5$, and $\epsilon = 2$. The gray shaded region is only accessible if the dilaton mass is fine-tuned according to Eq.~\eqref{eq_naturalness_bound_2}. The burgundy line indicates the parameter space accessible without a nonminimal dilaton-inflaton coupling. We show constraints from fifth-force modifications of the gravitational inverse square law~\textit{(red shaded region)}~\cite{Adelberger:2009zz,Murata:2014nra,Chen:2014oda,Lee:2020zjt} and from structure formation~\textit{(orange shaded region)}, where we demand that $T_\osc\gtrsim \keV$ [see Eq.~\eqref{eq_structure_formation_bound}]. We also include projected sensitivities, adapted from~\cite{Antypas:2022asj,AxionLimits}, for an optical cavity-cavity comparison experiment~\textit{(dashed brown line)}~\cite{Geraci:2018fax}, for the atom interferometry experiments AION~\textit{(dashed green line)}~\cite{Badurina:2021rgt} and MAGIS~\textit{(dashed blue line)}~\cite{MAGIS-100:2021etm}, for SrOH molecular clock spectroscopy~\textit{(dashed purple line)}~\cite{Kozyryev:2018pcp}, and for a variety of breathing-mode mechanical resonators~\cite{Manley:2019vxy}: a helium bar resonator~\textit{(cyan)}~\cite{Lorenzo:2016isc}, a sapphire cylindrical test mass~\textit{(green)}~\cite{Rowan:2000wd}, a micropillar resonator~\textit{(blue)}~\cite{pillarcitation}, and a quartz bulk acoustic wave resonator~\textit{(purple)}~\cite{bawcitation1,bawcitation2,bawcitation3}. }
		\label{fig:exclusion2}
	\end{figure}
 
\begin{figure}
		\centering
		\includegraphics[width=0.8\textwidth]{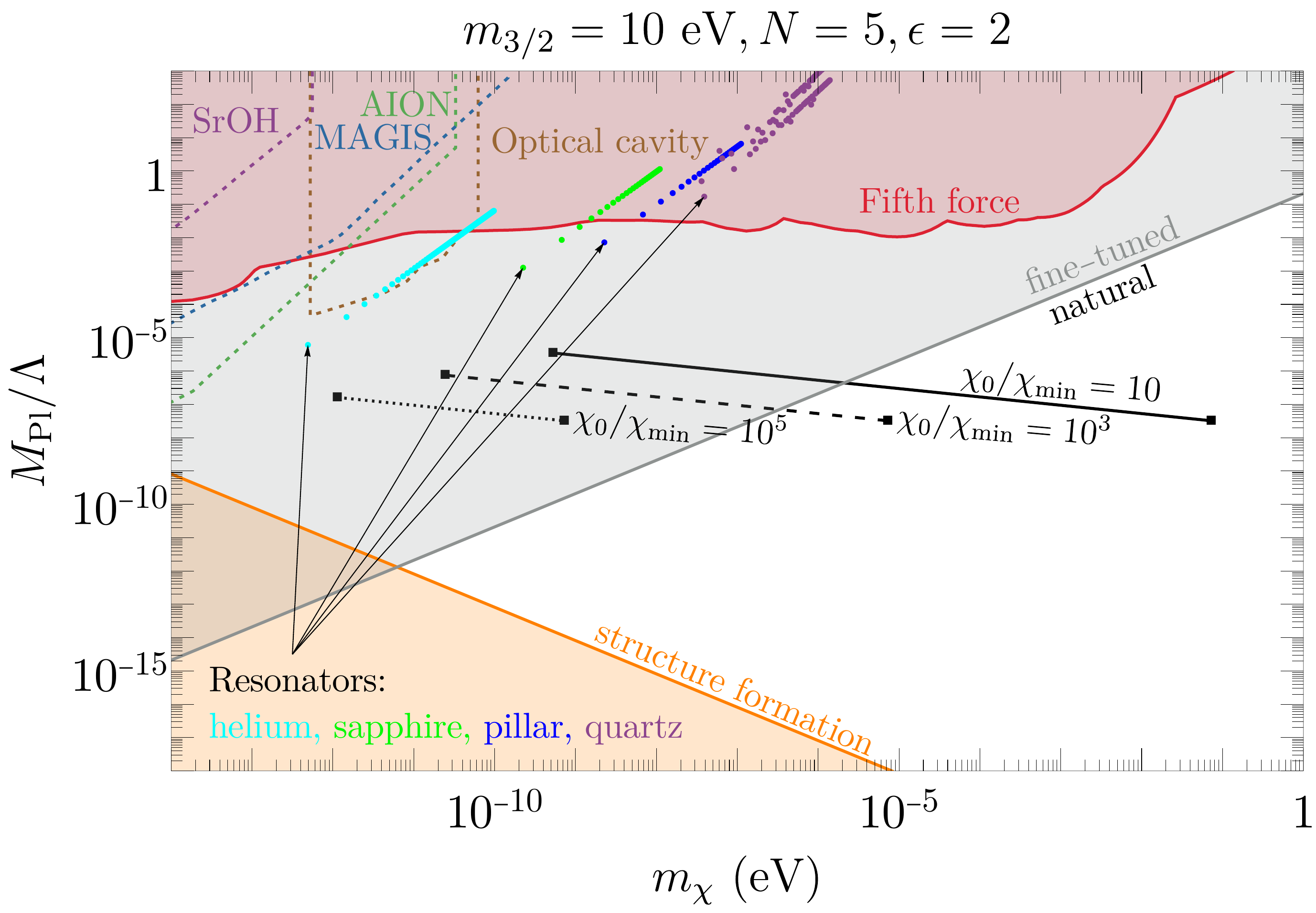}
		\caption{Exclusion plot for our ultralight dilaton, but with the initial condition $\chi_0 \gg \chimin$, corresponding to the scenario in Sec.~\ref{sec_misalignment_initial_k}. The black lines show the dilaton parameter space for different choices of the initial condition: $\chi_0/\chimin = 10$~\textit{(solid black line)}, $\chi_0/\chimin = 10^{5}$~\textit{(dashed black line)}, and $\chi_0/\chimin = 10^{9}$~\textit{(dotted black line)}. We fix the gravitino mass $m_{3/2} = 10$~eV, the number of colors $N = 5$, and $\epsilon = 2$. The bounds are the same as in Fig.~\ref{fig:exclusion2}. For the structure formation constraint $T_\osc\gtrsim \keV$, we use Eq.~\eqref{eq_structure_formation_bound2}.}
		\label{fig:exclusion}
	\end{figure}

	In Figs.~\ref{fig:exclusion2} and \ref{fig:exclusion} we show the possible mass $m_\chi$ and coupling strength $M_{\rm Pl}/\Lambda$ of the dilaton, assuming that it makes up all of the DM. For Fig.~\ref{fig:exclusion2} we assume that the initial condition for the dilaton is $\chi_0 \ll \chimin$, corresponding to the scenario discussed in Sec.~\ref{sec_misalignment_initial_origin}. We fix $\epsilon = 2$ and show the parameter space that yields the observed DM relic abundance for benchmark values $\chi_0/\chi_{\rm min} = 10^{-1}, 10^{-5}, 10^{-9}$. The requirement that $\overline k$ is not larger than the Planck scale leads to a lower bound on the dilaton mass, which is why the line for $\chi_0/\chi_{\rm min} = 10^{-1}$ is cut off at $m_\chi \sim 10^{-10}$~eV. The value of this bound depends on the gravitino mass and the number of colors; for our plot we fix $m_{3/2} = 10$~eV and $N = 5$.  Recall from Sec.~\ref{sec_misalignement} that in the minimal scenario where the inflaton does not couple to the dilaton, the initial condition is bounded as $\overline{\chi}_0 \lesssim 10^5$~eV. Fig.~\ref{fig:exclusion2} shows that most of the parameter space is inaccessible in this scenario. 
 
    In Fig.~\ref{fig:exclusion} we take the initial condition $\chi_0 \gg \chimin$, corresponding to the scenario discussed in Sec.~\ref{sec_misalignment_initial_k}. We again fix $\epsilon = 2$ and show the accessible parameter space for benchmark points $\chi_0/\chi_{\rm min} = 10, 10^3, 10^5$. We obtain upper and lower bounds on the dilaton mass from the requirement $\overline{\chi}_0 < \overline{k} < \sqrt{6} M_{\rm Pl}$.
	
	The naturalness bound in Eq.~\eqref{eq_naturalness_bound_1} translates into an upper bound on the coupling (equivalently, a lower bound on the dilaton mass) given by
	\begin{equation}\label{eq_naturalness_bound_2}
        \frac{M_\pl}{\Lambda}\lesssim \frac{2 \pi}{3} \frac{m_\chi}{m_{3/2}}
	\end{equation}
    The region of parameter space in tension with this bound is depicted by the gray areas in Figs.~\ref{fig:exclusion2}~and~\ref{fig:exclusion}, fixing $m_{3/2} = 10$~eV.
	
	The dilaton has no effect on tests of the equivalence principle because it couples to the trace of the entire stress-energy tensor~\cite{Damour:2010rp}. Instead, the main experimental constraint comes from searches for fifth-force modifications of the inverse square law~\cite{Adelberger:2009zz,Murata:2014nra,Chen:2014oda,Lee:2020zjt}. We also require the dilaton to behave like cold dark matter at $T > $~keV, so as not to affect the matter power spectrum [see Eqs.~\eqref{eq_structure_formation_bound} and~\eqref{eq_structure_formation_bound2}]. We remark that the relic density scales as $m_\chi^2/\Lambda^2 t_{\rm osc}^{3/2}$, where $t_{\rm osc}$ is the time at which the DM fields begin harmonic oscillations. The matter power spectrum effectively constrains $t_{\rm osc}$, which is why the bound in Fig.~\ref{fig:exclusion2} and Fig.~\ref{fig:exclusion} lies along a line of constant $m_\chi/\Lambda$.
 
    We include experimental projections, adapted from~\cite{Antypas:2022asj,AxionLimits}, for an optical cavity experiment~\cite{Geraci:2018fax}, atom interferometers~\cite{Badurina:2021rgt,MAGIS-100:2021etm}, SrOH molecular clock spectroscopy~\cite{Kozyryev:2018pcp}, and a variety of mechanical resonators~\cite{Manley:2019vxy,Lorenzo:2016isc,Rowan:2000wd,pillarcitation,bawcitation1,bawcitation2,bawcitation3}. These could further probe the parameter space for dilaton masses in the $10^{-14}$--$10^{-8}$~eV range. However, none of these experiments would probe parameter space that is not already in tension with the naturalness bound, Eq.~\eqref{eq_naturalness_bound_2}.

    Refs.~\cite{Cyncynates:2024bxw,Cyncynates:2024ufu} studied the impact of thermal corrections on the scalar DM potential due to the SM bath, and have shown that for sufficiently strong DM-SM interactions, these can displace the scalar from the minimum and lead to additional DM particle production through misalignment. Their analysis suggests that within the natural region in the mass range of $m_\chi\approx10^{-4}-1\eV$ and coupling strengths $d_{m_e}=\sqrt 2M_\pl/\Lambda \approx 10^{-6}-10^{-5}$, the thermal contribution to the dilaton potential could reproduce the measured DM abundance. We have neglected thermal effects on our model and leave a detailed study of this production channel to future work.
	
	The hallmark of the dilaton is that it couples to the trace of the stress-energy tensor as in Eq.~\eqref{eq:dilaton_couplings_general}. This universal coupling has some distinctive phenomenological consequences that are worth emphasizing. First, as stated previously, the dilaton does not lead to violations of the equivalence principle, but still induces a Yukawa-like modification to the Newtonian gravitational potential. Moreover, the couplings of the dilaton to the different SM fields are correlated. For example, if the dilaton were discovered by an experiment sensitive to oscillations of the electron mass, one could extract the dilaton mass and coupling strength, and then use them to make a prediction for the oscillation of the fine-structure constant. These effects --- the modification of the inverse square law without violating the equivalence principle, and the correlated couplings to different SM fields --- allow our ultralight dilaton to be distinguished from a generic scalar.
	
	\subsection{Constraints for the R-axion and the Dilatino}\label{subsec_r-axion_dilatino_pheno}
	
	Finally, we comment on the phenomenology of the R-axion $\vartheta$ and the dilatino $\tilde\omega$. In App.~\ref{app_SUGRA_R-axion_dilatino} we determine their couplings to the MSSM by applying a supersymmetric version of the Weyl rescaling argument that we used in Sec.~\ref{sec_dilaton_phenomenology} for the dilaton couplings. As expected from SUSY, the leading interactions are dimension-five operators suppressed by the same trans-Planckian scale $\Lambda$ as the dilaton interactions with the MSSM [cf. Eqs.~\eqref{eq_SUGRA_R-axion_MSSM_couplings} and~\eqref{eq_SUGRA_dilatino_MSSM_couplings}]. This implies that the R-axion and the dilatino are practically invisible in all terrestrial experiments. Even for the smallest possible suppression scale ($\Lambda = \sqrt{6} M_{\rm Pl}$) the R-axion couplings are several orders of magnitude below the existing and projected limits of the axion-electron, axion-proton, and axion-neutron couplings~\cite{AxionLimits}. It is also evident that all the modulus components cannot thermalize with the MSSM during the cosmological history of the Universe, and will not be produced in significant quantities by MSSM decays.

	Gravitino decays can produce relativistic dilatinos due to their much heavier mass, $m_{3/2}\gg m_{\tilde\omega}$ [cf. Eq.~\eqref{eq_CFT_Raxion_dilatino_masses}]. We assume that the dilaton and the R-axion constitute the majority of DM and that, regardless of how many gravitinos have been produced by the hidden sector dynamics, they make up only a small fraction of the energy density of DM. Consequently, the energy density of the warm dilatinos that are produced by gravitino decays must be small, in accordance with current bounds on warm DM~\cite{Irsic:2017ixq,Palanque-Delabrouille:2019iyz,Garzilli:2019qki,Yeche:2017upn}. 
	
	Furthermore, gravitino decays to dilatinos are rare because their coupling is suppressed; this coupling is determined by the dilatino's interaction with the goldstino $\widetilde G$ (the Nambu-Goldstone mode associated with broken SUSY). The gravitino gets its mass by eating the goldstino via the super-Higgs mechanism. The goldstino couples to the fermion component $\tilde\O$ of the operator in the UV, which we can parameterize by setting the spurion-fermion component to $\tilde\Sigma=\tilde G/M_\pl$ in Sec.~\ref{sec_SCFT_dilaton_stabilization}.\footnote{We obtained this normalization of $\tilde\Sigma$ because we set $\Sigma$ dimensionless, see the paragraph above Eq.~\eqref{eq_CFT_low_energy_SUSY_modulus_EFT}.} By plugging in $\tilde\Sigma$ into the effective Lagrangian in Eq.~\eqref{eq_CFT_low_energy_SUSY_modulus_EFT} we obtain a goldstino-dilatino-dilaton interaction 
	\begin{equation}
		\L_{\widetilde G-\widetilde\omega}\sim\frac{1}{2} \frac{m_{3/2}}{M_\pl}\left(\frac{\chimin}{k}\right)^{2\epsilon} \omega \tilde\omega \tilde G +\hc
	\end{equation}
	The goldstino couples very weakly to the dilatino, so gravitino decays are extremely rare.
	
	\section{Conclusions}

    In this paper, we presented a model of ultralight dilaton DM. The main challenge is that the dilaton mass and the IR scale are naturally of the same magnitude, whereas cosmological observations require a large hierarchy between the mass and the field oscillation amplitude. We circumvented this issue by constructing a supersymmetric dilaton model in which this hierarchy is protected by SUSY without fine-tuning. We then studied the production of dilaton DM through misalignment, where we found that the noncompact field range of the dilaton and the anharmonic shape of its potential significantly alter the final abundance. Although there is a wide range of viable parameter space for our model, to the best of our knowledge, no existing or proposed experiments can probe it.

    The spontaneous breaking of scale invariance in our SCFT is tied to SUSY-breaking: when SUSY is conserved the dilaton VEV vanishes and the CFT is unbroken, but when SUSY-breaking effects in the UV are taken into account, a nonzero IR scale is generated. This construction was necessary to control the SUSY-breaking corrections to the dilaton mass, which would otherwise make it too heavy to be a viable ULDM candidate. This led to an unusual stabilization mechanism where the conformal breaking is triggered by an irrelevant operator, and an exponentially small coupling $\lambda$ on the IR brane generates the UV/IR hierarchy. The same small parameter also controls the dilaton mass/IR scale hierarchy. While $\lambda$ is protected by SUSY and thus technically natural, we have not presented a dynamical way to generate a small $\lambda$.

    Although the low-energy dilaton EFT appears simple, our underlying UV model is quite involved, requiring SUSY in addition to a CFT (or extra dimensions in the holographic dual), as well as an elaborate stabilization mechanism that is far from generic. Another drawback in our model is that most of the parameter space is only accessible with nonminimal dilaton-inflaton couplings which require additional model-building. While a consistent model is possible without fine-tuning, it ends up somewhat contrived and difficult to probe experimentally, disfavoring the possibility of ultralight dilatonic DM.

    \acknowledgments

    We thank Keegan Humphrey, Yuri Shirman and Raman Sundrum for useful and enlightening discussions. The work of AB is supported by the National Science Foundation under grant number PHY2210361 and the Maryland Center for Fundamental Physics. CC is supported in part by the NSF grant PHY-2309456 and in part by the US-Israeli BSF grant 2016153. MG and ZHA are supported by the Israel Science Foundation under Grant No. 1302/19 and 1424/23. MG is also supported by the US-Israeli BSF grant 2018236 and the NSF-BSF grant 2021779. AI is supported by a Mafalda and Reinhard Oehme Postdoctoral Research Fellowship from the Enrico Fermi Institute at the University of Chicago. We also thank the Munich Institute for Astro-, Particle and BioPhysics (funded by the DFG under Germany's Excellence Strategy EXC-2094-390783311) for its hospitality while this paper was being concluded.
	
	\appendix
	
	\section{The Low-energy Modulus EFT from the Warped 5D Picture}\label{app_5D_RS_EFT_proof}
	
	In this appendix we derive the low energy modulus EFT
	\begin{equation}\label{eq_5D_low_energy_SUSY_modulus_EFT}
		\L_\eff=\int \odif[order=4]{\theta}
        \left[\frac{3N^2}{4\pi^2}\bs\omega^\dagger\bs\omega-\frac{\bs\omega^\dagger\bs\omega}{1+\epsilon}\left|\bs\Sigma\left(\frac{\bs\omega}{k}\right)^\epsilon\right|^2\right]
        +\int\odif[order=2]{\theta}\bs\omega^3\left[\kappa+\lambda\bs\Sigma\left(\frac{\bs\omega}{k}\right)^\epsilon\right]+\hc
	\end{equation}
	The derivation assumes that SUSY-breaking effects and the parameter $\lambda$  are both small, which allows us to neglect corrections to the formula above. This set of assumptions is consistent with the SCFT model described in the main text in Sec.~\ref{sec_SCFT}, where we make use of this EFT to analyze the dynamics of the modulus field in low energies. We utilize the AdS/CFT correspondence to derive this EFT by calculating the effective Lagrangian of the supersymmetric radion field in the dual supersymmetric RS model.
	
	\subsection{SUSY in the RS Background}
	
	We work in the RS background on an $S^1/\mathbb{Z}_2$  orbifold, using the circle coordinates denoted by $y\in[0,2\pi]$ for the compactified dimension. The UV and the IR branes are identified with the orbifold fixed points $y=0,\pi$, respectively. The metric is given by
	\begin{equation}\label{eq_5D_RS_background_metric}
		\odif{s}^2=e^{-2\sigma r}\eta_{\mu\nu}\odif{x}^\mu\odif{x}^\nu-r^2\odif{y}^2,
	\end{equation}
	where $r$ is the radius (the radion) of the extra dimension, $\sigma=k|y|$ and $k$ is the AdS curvature. This metric solves the Einstein equations when the bulk CC is $\Lambda_b=6M_5^3k^2$ and the brane tensions are tuned to $\Lambda_0=-\Lambda_\pi=\Lambda_b/k$, where $M_5$ is the 5D Planck scale.\footnote{In our convention  the 5D Einstein-Hilbert Lagrangian is $-\frac{1}{2}M_5^3\sqrt{g}\R_5$.}
	
	The supersymmetric extension of the RS model was presented in~\cite{Marti:2001iw} using $\N=1$ superfields. The radion is promoted to a chiral superfield
	\begin{equation}
		\bs T=r+ib+\sqrt{2}\theta\Psi_T +\theta^2F_T,
	\end{equation}
	where $b$ is the 5th component of the graviphoton, $\Psi_T$ is the 5th component of the right handed gravitino and $F_T$ is a complex auxiliary field.  
	
	We introduce a bulk hypermultiplet $(\bs\Phi,\bs\Phi^c)$ (written as two $\N=1$ chiral superfields with orbifold parity assignments $(+,-)$) that is dual to the operator $\bs\O$ in the SCFT~\cite{Cacciapaglia:2008bi}. Its mass is parameterized by the dimensionless number $c$~\cite{Gherghetta:2000qt}, which is related to the anomalous dimension $\epsilon$ of the operator $\O$ via $c=\frac{3}{2}+\epsilon$ (for $c>1/2$, which we assume throughout). The matter fields generate a potential for the radion that stabilizes it at a finite value, which is interpreted in the SCFT as the generation of the IR mass gap by the weak deformation in the UV. This is the supersymmetric version of the Goldberger-Wise mechanism~\cite{Goldberger:1999uk} (see also~\cite{Maru:2003mq}).
	
	The 5D Lagrangian of the bulk matter fields is given by~\cite{Marti:2001iw}
	\begin{equation}
		\begin{aligned}
			\L_5^{\rm m}&=\int \odif[order=4]{\theta} \frac{1}{2}\big( \bs T+\bs T^\dagger \big)e^{-\sigma(\bs T+\bs T^\dagger)}\left(\bs\Phi^\dagger\bs\Phi+\bs\Phi^c\bs\Phi^{c\dagger}\right)\\
			&+\int\odif[order=2]{\theta}e^{-3\bs T\sigma}\left[\frac{1}{2}\bs\Phi^c\dvec\partial_y\bs\Phi+c\bs T\sigma'\bs\Phi^c\bs\Phi+W_0(\bs\Phi)\delta(y)+W_\pi(\bs\Phi)\delta(y-\pi)\right]+\hc
		\end{aligned}
	\end{equation}
	$\dvec\partial_y$ is the double-sided derivative and $W_{0,\pi}$ are the brane superpotentials, which are functions of the even $\bs\Phi$ and possibly other brane-localized fields. The brane-localized superpotentials cause the bulk superfield $(\bs\Phi,\bs\Phi^c)$ to obtain VEVs on the branes, similar to the brane-localized potentials in the usual Goldberger-Wise mechanism~\cite{Goldberger:1999uk}.
    After integrating this Lagrangian over superspace, the Lagrangian of the on-shell chiral superfield in the RS background is recovered~\cite{Gherghetta:2000qt}.
    The $e^{-\sigma\bs T}$ warp factors in the K\"ahler potential and in the superpotential should be understood as a rescaling of the conformal compensator, which is equivalent to the rescaling of the metric in the AdS solution in Eq.~\eqref{eq_5D_RS_background_metric}.
	
	It is convenient to introduce rescaled matter fields whose zero-mode profile will be flat across the extra dimension,
	\begin{equation}
		\hat{\bs\Phi}\equiv\bs\Phi e^{(c-\frac{3}{2})\sigma\bs T},\qquad \hat{\bs\Phi}^c\equiv\bs\Phi^c e^{(-c-\frac{3}{2})\sigma\bs T}.
	\end{equation}
	This eliminates the warp factor and the mass term in the bulk superpotential in the matter Lagrangian, which is now
	\begin{equation}\label{eq_5D_matter_lagrangian}
		\begin{aligned}
			\L_5^{\rm m}&=\int \odif[order=4]{\theta} \frac{1}{2}\big( \bs T+\bs T^\dagger \big) \left[\hat{\bs\Phi}^\dagger\hat{\bs\Phi} e^{(\frac{1}{2}-c)\sigma(\bs T+\bs T^\dagger)}+\hat{\bs\Phi}^c\hat{\bs\Phi}^{c\dagger}e^{(\frac{1}{2}+c)\sigma(\bs T+\bs T^\dagger)}\right]\\
			&+\int\odif[order=2]{\theta}\left[\frac{1}{2}\hat{\bs\Phi}^c\dvec\partial_y\hat{\bs\Phi}+W_0\big(\hat{\bs\Phi}\big)\delta(y)+e^{-3\pi k \bs T}W_\pi \big(\hat{\bs\Phi} e^{(\frac{3}{2}-c)\pi k \bs T}\big)\delta(y-\pi)\right]+\hc
		\end{aligned}
	\end{equation}
	
	\subsection{Effective 4D Lagrangian for the Dilaton}\label{subapp_SUSYRS_effective_4D_theory}
	
	The VEV of the radion $\braket{r}$ determines the size of the extra dimension and hence the Kaluza-Klein (KK) scale. In the SCFT picture this corresponds to the IR scale where spontaneous breaking is triggered, which is determined by the dilaton VEV $\braket\chi$. At energies below the KK scale the dynamics in the fifth dimension can be integrated out, and the relevant physics is captured by an effective 4D Lagrangian. For the supersymmetric radion $\bs T$, the effective Lagrangian after dimensional reduction is given by~\cite{Luty:2000ec}
	\begin{equation}\label{eq_5D_general_superspace_dilaton_Lagrangian}
		\L_\eff=\frac{3M_5^3}{k}\int \odif[order=4]{\theta}\left(e^{-\pi k(\bs T+\bs T^\dagger)}-1\right)+\L_\eff^{\rm m}
		=\frac{3M_5^3}{k^3}\int \odif[order=4]{\theta}(\bs\omega^\dagger\bs\omega-k^2)+\L_\eff^{\rm m}.
	\end{equation}
	In the second equality we defined the superfield $\bs\omega$, which we identify as the modulus of the dual SCFT,  
	\begin{equation}
		\bs\omega\equiv k e^{-\pi k \bs{T}}.
	\end{equation}
	
	The K\"ahler potential in Eq.~\eqref{eq_5D_general_superspace_dilaton_Lagrangian} originates from the dimensional reduction of the pure 5D SUGRA Lagrangian in the RS background~\cite{Luty:2000ec}. We obtain the kinetic term $\bs\omega^\dagger\bs\omega$ of the radion/modulus superfield, the same way it is found in non-supersymmetric RS~\cite{Goldberger:1999un}. The extra $-1$ factor, which would vanish after integrating over the rigid superspace coordinates used throughout this paper, provides the correct coefficient of the Ricci term $M_5^3/2k(1-e^{-2\pi k r})$ in RS~\cite{Randall:1999ee} when we add the appropriate measure to $\odif[order=4]{\theta}$ in SUGRA (see App.~\ref{app_SUGRA_R-axion_dilatino}). Using the holographic relation $N^2=4\pi^2(M_5/k)^3$~\cite{Gubser:1999vj}, we see that this modulus kinetic term matches the first term of the K\"ahler potential in Eq.~\eqref{eq_5D_low_energy_SUSY_modulus_EFT}.
	
	The second $\L_\eff^{\rm m}$ term in Eq.~\eqref{eq_5D_general_superspace_dilaton_Lagrangian}  is the effective 4D Lagrangian obtained from additional matter in the bulk, whose backreaction on the RS metric in Eq.~\eqref{eq_5D_RS_background_metric} is assumed to be small. It is clear that without this bulk contribution in Eq.~\eqref{eq_5D_general_superspace_dilaton_Lagrangian}, the modulus field is massless and the radion VEV is undetermined (corresponding to a purely spontaneously broken SCFT). The bulk contribution to the effective Lagrangian is obtained by plugging in the VEV profiles of the matter fields into the 5D bulk matter Lagrangian $\L_5^{\rm m}$ and integrating over the fifth dimension~\cite{Csaki:2023pwy}.
	
	In our case, we introduce a chiral superfield $(\bs\Phi,\bs\Phi^c)$ in the bulk to stabilize the radion. The 5D Lagrangian of the bulk matter is given in Eq.~\eqref{eq_5D_matter_lagrangian}. The VEV profiles of $\bs\Phi$ and $\bs\Phi^c$ are the zero-mode ($x^\mu$-independent) solutions of the equations of motion (EOM). We obtain the EOM by varying $\L_5^{\rm m}$ with respect to $\hat{\bs\Phi}$ and $\hat{\bs\Phi}^c$ and fixing the modulus components to their VEVs, and these turn out to be a pair of coupled first-order differential equations due to the K\"ahler potential in Eq.~\eqref{eq_5D_matter_lagrangian}. These are difficult to solve in general, but under our assumption that SUSY-breaking effects are small (i.e., the SCFT is nearly supersymmetric), we can set $\braket{F_T}=0$ and ignore the $F_\Phi$ and $F_\Phi^c$ terms arising from the K\"ahler potential. In total, the simplified EOMs do not involve the K\"ahler potential and are found to be
	\begin{equation}\label{eq_5D_simplified_zero-mode_EOM}
		\partial_y\hat{\bs\Phi}=0,\qquad
		\partial_y\hat{\bs\Phi}^c=\pdv{W_0}{\hat{\bs\Phi}}\delta(y)+e^{-3\pi k \bs T}\pdv{W_\pi}{\hat{\bs\Phi}}\delta(y-\pi).
	\end{equation}
	The $\hat{\bs\Phi},\hat{\bs\Phi}^c$ bulk solutions (i.e. the solution to the EOM where the singular terms are ignored) are constant, corresponding to the zero-mode bulk profiles
	\begin{equation}\label{eq_5D_zero-mode_profiles}
		\bs\Phi(y)=\bs\Phi_0e^{(\frac{3}{2}-c)\sigma\bs T},\qquad \bs\Phi^c(y)=\bs\Phi^c_0e^{(\frac{3}{2}+c)\sigma\bs T}.
	\end{equation}
	All the components of $\bs\Phi,\bs\Phi^c$ have exactly the same wave function across the extra dimension, as expected in the SUSY-conserving limit. By matching the discontinuities in the bulk $\bs\Phi^c$ solution with the singular terms induced by the brane superpotentials $W_0,W_\pi$,  we find that the zero-mode boundary values are fixed by the jump equations
	\begin{equation}\label{eq_5D_jump_equations}
		\bs\Phi^c_0=\frac{1}{2}\pdv{W_0}{\bs\Phi}\bigg|_{y=0}
        =-\frac{1}{2}\pdv{W_\pi}{\bs\Phi}\bigg|_{y=\pi}e^{-(\frac{3}{2}+c)\pi k \bs T}.
	\end{equation}
    This is the supersymmetric version of the boundary conditions used in the Goldberger-Wise construction to fix the unknown coefficients of the bulk solution~\cite{Goldberger:1999uk}.
	
	Observe that a SUSY-breaking VEV for $F_\Phi$ on the UV brane leads to nonvanishing $F_\Phi,F_\Phi^c\neq0$ profiles in the bulk, which in turn source corrections to the zero-mode profiles of the scalar components $\Phi,\Phi^c$ through the K\"ahler potential terms we neglected in Eq.~\eqref{eq_5D_simplified_zero-mode_EOM}. The inclusion of these corrections will result in different profiles for the different components of $\bs\Phi$ and $\bs\Phi^c$ due to SUSY-breaking effects~\cite{Goh:2003yr}. In the small SUSY-breaking limit we consider, we do not need to find these corrections explicitly.
	
	Now we plug in the zero-mode profiles $\hat{\bs\Phi}$ and $\hat{\bs\Phi}^c$ into $\L_5^{\rm m}$ in Eq.~\eqref{eq_5D_matter_lagrangian}. The bulk superpotential piece $\hat{\bs\Phi}^c\dvec\partial_y\hat{\bs\Phi}$ vanishes,\footnote{When plugging in the zero-mode profile in $\partial_y\hat{\bs\Phi}^c$, one must carefully include the contribution of the singular pieces on the branes to see that it vanishes. It is easier to integrate by parts and replace the double-sided derivative by $\hat{\bs\Phi}^c\partial_y\hat{\bs\Phi}$, which has no ambiguity about the singular terms on the branes, since in the orbifold integrals of total derivatives are always zero.} and we are left with the K\"ahler potential and the brane superpotential contributions
	\begin{equation}
		\begin{aligned}
			\L_{5,\text{0-mode}}^{\rm m}&=\int \odif[order=4]{\theta} \frac{1}{2}\big( \bs T+\bs T^\dagger \big) \left[\hat{\bs\Phi}^\dagger\hat{\bs\Phi} e^{(\frac{1}{2}-c)\sigma(\bs T+\bs T^\dagger)}+\hat{\bs\Phi}^c\hat{\bs\Phi}^{c\dagger}e^{(\frac{1}{2}+c)\sigma(\bs T+\bs T^\dagger)}\right]\\
			&+\int\odif[order=2]{\theta}\left[W_0\big(\hat{\bs\Phi}\big)\delta(y)+e^{-3\pi k \bs T}W_\pi \big(\hat{\bs\Phi} e^{(\frac{3}{2}-c)\pi k \bs T}\big)\delta(y-\pi)\right]+\hc
		\end{aligned}
	\end{equation}
	Next we integrate this expression over the extra dimension to obtain the effective 4D Lagrangian of the bulk matter fields and express it in terms of the modulus field $\bs\omega$,
	\begin{equation}\label{eq_5D_general_effective_4D_bulk_matter_Lagrangian}
		\begin{aligned}
			\L_\eff^{\rm m}&=\int \odif[order=4]{\theta}\left[\bs\Phi^\dagger_0\bs\Phi_0 \frac{(\bs\omega^\dagger\bs\omega/k^2)^{c-\frac{1}{2}}-1}{(\frac{1}{2}-c)k}+\bs\Phi^c_0\bs\Phi^{c\dagger}_0\frac{(\bs\omega^\dagger\bs\omega/k^2)^{-c-\frac{1}{2}}-1}{(\frac{1}{2}+c)k}\right]\\
			&+\int\odif[order=2]{\theta}\left[W_0(\bs\Phi_0)+(\bs\omega/k)^3W_\pi\big(\bs\Phi_0(\bs\omega/k)^{c-\frac{3}{2}}\big)\right]+\hc
		\end{aligned}
	\end{equation}
	Eqs.~\eqref{eq_5D_general_effective_4D_bulk_matter_Lagrangian} and~\eqref{eq_5D_jump_equations} determine the bulk matter contribution to the effective Lagrangian of the modulus below the KK scale.   
	
	\subsection{An appropriate choice of Brane Superpotentials for our Model}
	
	Here we choose specific brane superpotentials that will generate the effective modulus Lagrangian we analyzed in Sec.~\ref{sec_SCFT}. On the UV brane a quadratic superpotential will give $\bs\Phi$ a VEV, and on the IR brane we put a constant term and a small tadpole,
	\begin{equation}
		W_0=\alpha(\tfrac{1}{2}\bs\Phi^2-\bs\Sigma\bs\Phi k^{3/2}),\qquad W_\pi =\kappa k^3+\lambda\bs\Phi k^{3/2}.
	\end{equation}
	$\bs\Sigma$ and all the coefficients are dimensionless. $\bs\Sigma$ on the UV brane corresponds to the spurion couplings of $\bs\O$ in the equivalent SCFT [cf. Eq.~\eqref{eq_CFT_weak_SCFT_deformation}], and its auxiliary term $F_\Sigma$ is assumed to be small so that the theory remains nearly supersymmetric. The constant term $\kappa k^3$ in the IR superpotential is a mistuning of the IR brane tension, and accordingly will generate a scale-invariant quartic term $\chi^4$ in the dilaton potential. $\kappa$ and $\lambda$ in the IR superpotential can be taken to be zero or extremely small by virtue of the nonrenormalization theorem. 
	
	We solve the supersymmetric jump equations in Eq.~\eqref{eq_5D_jump_equations} to find the UV initial conditions,
	\begin{equation}
		\bs\Phi_0=\bs\Sigma k^{3/2}-\frac{\lambda}{\alpha}\left(\frac{\bs\omega}{k}\right)^c\bs\omega^{3/2},\qquad \bs\Phi^c_0=-\frac{\lambda}{2}\left(\frac{\bs\omega}{k}\right)^c\bs\omega^{3/2}.
	\end{equation}
	When $\lambda$ is small and the bulk mass $c$ is fairly large, the odd field vanishes $(\bs\Phi^c\equiv0)$ and $\bs\Phi_0=\bs\Sigma k^{3/2}$. Plugging these back into Eq.~\eqref{eq_5D_general_effective_4D_bulk_matter_Lagrangian}, the dominant contribution to the effective 4D Lagrangian for the modulus field is (disregarding terms independent of $\bs\omega$)
	\begin{equation}\label{eq_app_effective_matter_Lagrangian}
		\L_\eff^{\rm m}=-k^2\int \odif[order=4]{\theta}\frac{\bs\Sigma^\dagger\bs\Sigma}{c-\frac{1}{2}}\left(\frac{\bs\omega^\dagger\bs\omega}{k^2}\right)^{c-\frac{1}{2}}+\int\odif[order=2]{\theta}\bs\omega^3\left[\kappa+\lambda\bs\Sigma\left(\frac{\bs\omega}{k}\right)^{c-\frac{3}{2}}\right]+\hc+\ldots
	\end{equation}
	Recalling that the mass of the bulk field $(\bs\Phi,\bs\Phi^c)$ is related to the anomalous dimension of the corresponding operator $\bs\O$ via $c=\frac{3}{2}+\epsilon$, we obtain from the equation above the remaining interactions of the effective low energy modulus Lagrangian in Eq.~\eqref{eq_5D_low_energy_SUSY_modulus_EFT}.
	
	The additional terms we disregarded above are explicitly  
	\begin{equation}
		\begin{aligned}
			\ldots &=k^2|\lambda|^2\int \odif[order=4]{\theta}\left(\frac{\bs\omega^\dagger\bs\omega}{k^2}\right)^{c+\frac{3}{2}}
			\left[\frac{1}{|\alpha|^2}\frac{(\bs\omega^\dagger\bs\omega/k^2)^{c-\frac{1}{2}}-1}{\frac{1}{2}-c}
				+\frac{1}{4}\frac{(\bs\omega^\dagger\bs\omega/k^2)^{-c-\frac{1}{2}}-1}{\frac{1}{2}+c}\right] \\
			&-k^2\frac{\lambda}{\alpha}\int\odif[order=4]{\theta} \bs\Sigma^\dagger \left(\frac{\bs\omega}{k}\right)^{c+\frac{3}{2}}\frac{(\bs\omega^\dagger\bs\omega/k^2)^{c-\frac{1}{2}}-1}{\frac{1}{2}-c}-\frac{\lambda^2}{2\alpha}\int\odif[order=2]{\theta}\bs\omega^3\left(\frac{\bs\omega}{k}\right)^{2c}+\hc
		\end{aligned}
	\end{equation}
    The first line gives $O(\lambda^2)$ corrections to the K\"ahler potential, which we neglect because $\lambda$ is small. Similarly, we ignore both terms in the second line because they are suppressed by $\sim\lambda(\bs\omega/k)^{c+\frac32}$ w.r.t. the K\"ahler potential and the superpotential in Eq.~\eqref{eq_app_effective_matter_Lagrangian}, respectively.
	
	\section{Dilaton Energy Density Evolution}\label{app_energy_density}
	
	In this appendix we study the time evolution of the DM fields and their energy density to derive Eq.~\eqref{eq:energydensity}. Our starting point is the effective Lagrangian Eq.~\eqref{eq:lagrangian}. Assuming spatially homogeneous fields $\phi = \phi(t), \vartheta = \vartheta(t)$, if we measure time in units of $1/m_\chi$ ($t \rightarrow t/m_\chi$), the Lagrangian takes the simple form
	\begin{equation}
		\frac{\L_\eff}{m_\chi^2 f^2} = \frac{1}{2} \dot{\phi}^2+ \frac{1}{2} \phi^2 \dot{\vartheta}^2  \underbrace{-\frac{1}{\epsilon-1}\left(\frac{\phi^{2+2\epsilon}}{2+2\epsilon}-\frac{\phi^{3+\epsilon}}{3+\epsilon}\cos[(3+\epsilon)\vartheta]\right)+\frac{1}{2(\epsilon^2+4\epsilon+3)}}_{-V(\phi,\vartheta)},
	\end{equation}
    where $V(\phi,\vartheta)$ is the potential normalized in units of $m_\chi^2f^2$. For the large $\phi$ under consideration we can neglect the constant term in the potential. The EOM during RD are
	\begin{equation}
			\ddot{\phi} + \frac{3}{2t} \dot{\phi} - \phi \dot{\vartheta}^2  + \pdv{V}{\phi} = 0 , \qquad
			\phi^2 \ddot{\vartheta} + \frac{3}{2t} \phi^2 \dot{\vartheta} + 2 \phi \dot{\phi} \dot{\vartheta} + \pdv{V}{\vartheta}  = 0.
	\end{equation}
	We are interested in the energy density $\rho$, which is given (in units of $m_\chi^2 f^2$) by
	\begin{equation}
		\rho = \frac{1}{2} \dot{\phi}^2 + \frac{1}{2} \phi^2 \dot{\vartheta}^2 + V(\phi, \vartheta) .
	\end{equation}
	
	For a large initial condition $\phi_0 \gg 1$, we approximate the effective potential as $V \approx \phi^{2+2\epsilon}/(2(\epsilon^2-1))$. Thus, the initial energy density is
	\begin{equation}
		\rho_{\rm initial} \approx \frac{\phi_0^{2+2\epsilon}}{2(\epsilon^2-1)}.
	\end{equation}
	Note that since the approximate potential is central (independent of $\vartheta$), the $\vartheta$ EOM is trivial. The approximate EOM for $\phi$ is then
	\begin{equation}
		0 = \ddot{\phi} + \frac{3}{2t} \dot{\phi} + \frac{1}{\epsilon - 1} \phi^{1+2\epsilon} .
	\end{equation}
	We can rewrite this equation in terms of rescaled variables $z \equiv \phi / \phi_0$ and $\tau \equiv t \phi_0^{\epsilon}$,
	\begin{equation}
		0 = \odv[order=2]{z}{\tau} + \frac{3}{2 \tau} \odv{z}{\tau} + \frac{1}{\epsilon -1} z^{1+2\epsilon} .
	\end{equation}
	This rescaling makes it manifest that the field at $z\approx 1$ is frozen by Hubble friction until a characteristic timescale $\tau_\anh \sim 3 (\epsilon-1)/2$, or equivalently until
	\begin{equation}\label{eq:t1}
		t_\anh = \frac{3(\epsilon-1)}{2} \phi_0^{-\epsilon} .
	\end{equation}
	So for $t < t_\anh$ we have $\rho(t) = \rho_{\rm initial}$.
	
	Once $t > t_\anh$, the DM undergoes complicated anharmonic oscillations, but we can identify a simple scaling behavior for the energy density. In the absence of Hubble friction, $\rho$ would be conserved. Using the equations of motion, we can write the time derivative of the energy density as
	\begin{equation}\label{eq:dissipation}
		\odv{\rho}{t}=-3H(\dot{\phi}^2 + \phi^2 \dot{\vartheta}^2)\implies\odv{\log\rho}{\log t} = -3 \frac{ \frac{1}{2}(\dot{\phi}^2 + \phi^2 \dot{\vartheta}^2 )}{\rho}.
	\end{equation}
	The fraction on the right hand side is the ratio of kinetic energy to total energy. To simplify this equation, we can average the right-hand side over one oscillation period. Again approximating the potential as a $\phi^{2+2\epsilon}$ power law, the virial theorem implies that the average fraction of kinetic energy is $(1+\epsilon)/(2+\epsilon)$. Eq.~\eqref{eq:dissipation} then implies a simple power-law scaling
	\begin{equation}\label{eq:energydensity1}
		\rho \approx \frac{\phi_0^{2+2\epsilon}}{2(\epsilon^2-1)} \left( \frac{t}{t_\anh} \right)^{-3(1+\epsilon)/(2+\epsilon)} \propto a^{-6(1+\epsilon)/(2+\epsilon)} .
	\end{equation}
	Note that in this regime the energy density is always redshifting faster than radiation, since $\epsilon > 1$.
	
	Eventually, the fields fall in toward the minimum at $\phi = 1, \vartheta = 0$. Once they have dissipated enough energy, the fields remain close enough to the minimum that the potential is well approximated by a quadratic function. By expanding the potential about the minimum, we find that this occurs when the energy density falls below $\rho = 1/(3+\epsilon)^2$, which according to  Eq.~\eqref{eq:energydensity1} occurs at the time $t_\osc$:
	\begin{equation}\label{eq:t2}
        t_\osc = \left[ \frac{(3+\epsilon)^2}{\epsilon^2-1}\phi_0^{2+2\epsilon} \right]^{(2+\epsilon)/(3+3\epsilon)} t_\anh.
	\end{equation}
	
	For $t > t_\osc$ the fields undergo harmonic oscillations about the minimum of the potential, with the energy density redshifting like $a^{-3}$ (cold  matter). Combining this with Eqs.~\eqref{eq:t1},~\eqref{eq:energydensity1}, and~\eqref{eq:t2} yields the piecewise expression for the energy density stated in Eq.~\eqref{eq:energydensity} in the main text.
	
	\section{R-axion and Dilatino Couplings to the MSSM}\label{app_SUGRA_R-axion_dilatino}
	
	In Sec.~\ref{sec_dilaton_phenomenology} we derived the dilaton-SM couplings by moving from the Jordan frame to the Einstein frame with a Weyl transformation $g_{\mu\nu}\rightarrow\Omega^2g_{\mu\nu}$ as written in Eq.~\eqref{eq_Weyl_transformation}. This procedure can be extended to the entire modulus supermultiplet in SUGRA, which we use here to derive the couplings of the R-axion and dilatino to the MSSM fields. We use the $\N=1$ superspace formulation of 4D SUGRA of~\cite{Wess:1992cp,RauschdeTraubenberg:2020kol}, where in the Jordan frame the Lagrangian is
	\begin{equation}\label{eq_SUGRA_lagrangian}
		\L_{\rm SUGRA}=\int\odif[order=4]{\theta}f(\bs\Phi_i,\bs\Phi_{\bar i}^\dagger)+\int\odif[order=2]{\theta}W(\bs\Phi_i)+\hc
	\end{equation}
	$\bs\Phi_i=\{\Phi_i,\psi_i,F_i\}$ are the chiral superfields of the theory ($\bs\Phi_{\bar i}^\dagger$ are their Hermitian conjugates), $f$ is the superspace kinetic energy and $W$ is the superpotential. We disregard gauge interactions for simplicity, since they are classically scale-invariant and thus unaffected by a Weyl transformation. $\theta$ and $\bar\theta$ should be understood as ``curved'' superspace coordinates that transform under local Lorentz transformations like $x^\mu$ in GR, and the gravity supermultiplet enters through super-determinants in their measure (implicit in our notation) like $\sqrt{-g}$ in GR.
    
    The complete calculation of the component-level Lagrangian from Eq.~\eqref{eq_SUGRA_lagrangian} is explained in~\cite{Wess:1992cp,RauschdeTraubenberg:2020kol}. Here we highlight the important details relevant for our purposes; the Ricci term obtained from Eq.~\eqref{eq_SUGRA_lagrangian} is $\L_{\rm J}=-\frac{1}{6}f(\Phi_i,\Phi_{\bar i}^*)\sqrt{-g}\R_4$ which is noncanonical, and likewise the gravitino kinetic term is also noncanonical. To move to the Einstein frame, where $\L_{\rm EH}=-\frac{1}{2}M_\pl^2\sqrt{-g}\R_4$ and the gravitino is canonically normalized, one performs a Weyl transformation and a gravitino shift. In this frame the K\"ahler potential is
	\begin{equation}\label{eq_SUGRA_Kahler_potential}
		K=-3M_\pl^2 \ln[-f(\Phi_i,\Phi_{\bar i}^*)/3M_\pl^2].
	\end{equation}
    In our case, this noncanonical K\"ahler potential will include higher-order terms that induce interactions between the modulus and the MSSM fields. The novel R-axion and dilatino couplings are found by examining the SUGRA fermion Lagrangian, specifically the kinetic term and the 2-fermion interactions, given by
	\begin{equation}\label{eq_SUGRA_fermion_sector}
		\frac{1}{\sqrt{-g}}\L_{\psi}=
		\frac{i}{2}K_{i \bar i}(\psi^{\bar i \dagger}\bar\sigma^\mu \check{D}_\mu\psi^i-\check{D}_\mu\psi^{\bar i\dagger}\bar\sigma^{\mu}\psi^i)
		-\frac{1}{2}e^{K/2M_\pl^2}\D_i\D_jW\psi^i\psi^j+\hc,
	\end{equation}
	where the subscript $\Box_i$ $(\Box_{\bar i})$ is the derivative with respect to $\Phi_i$ $(\Phi^*_{\bar i})$. The fermion derivatives are covariant with respect to the K\"ahler manifold and its $\mathrm{U}(1)$-connection,
	\begin{equation}\label{eq_SUGRA_covaraint_fermion_derivative}
		\check{D}_\mu\psi^i=\partial_\mu\psi^i+\Gamma^i_{jk}\partial_\mu\Phi^j\psi^k-\frac{1}{4M_\pl^2}(K_j\partial_\mu\Phi^j-K_{\bar j}\partial_\mu\Phi^{\bar j*})\psi^i.
	\end{equation}
	$\Gamma^i_{jk}\equiv K^{i\bar i}K_{\bar i jk}$ is the Christoffel symbol of the K\"ahler manifold. The superpotential derivatives are defined as
	\begin{equation}
		\begin{aligned}
			\D_iW&=W_i+\frac{K_iW}{M_\pl},\\
			\D_i\D_jW&=W_{ij}+\frac{K_{ij}W}{M_\pl^2}+\frac{K_i\D_jW}{M_\pl^2}+\frac{K_j\D_iW}{M_\pl^2}-\frac{K_iK_jW}{M_\pl^4}-\Gamma^k_{ij}D_kW.
		\end{aligned}
	\end{equation}
    
    To determine the correct K\"ahler potential in our model, we start with the supersymmetric extension of the effective 4D Lagrangian in Eq.~\eqref{eq:eff4Dgravityaction},
	\begin{equation}
		\L_{\rm J}=\int\odif[order=4]{\theta}\left[\frac{3M_5^3}{k}\left(\frac{\bs\omega^\dagger\bs\omega}{k^2}-1\right)-3M_0^2+\bs\Phi^\dagger\bs\Phi\right]+\int\odif[order=2]{\theta}W(\bs\Phi)+\hc
	\end{equation}
	The first term is the effective 4D SUGRA Lagrangian obtained after integrating out the extra dimension in RS~\cite{Luty:2000ec}, and the second term is a localized Ricci term. $\bs\Phi$ stands for any UV-localized chiral superfield. According to Eq.~\eqref{eq_SUGRA_Kahler_potential}, the appropriate K\"ahler potential is
	\begin{equation}\label{eq_SUGRA_app_Kahler_potential}
		\begin{aligned}
			K&=-3M_\pl^2\ln\left[\frac{M_0^2}{M_\pl^2}+\frac{M_5^3}{kM_\pl^2}\left(1-\frac{\omega^*\omega}{k^2}\right)-\frac{\Phi^*\Phi}{3M_\pl^2}\right]
			\simeq \frac{3M_5^3}{k^3}\omega^*\omega+\Omega^2\Phi^*\Phi\\
			&\longrightarrow \omega^*\omega+\left(1+\frac{\omega^*\omega}{3M_\pl^2}\right)\Phi^*\Phi.
		\end{aligned}
	\end{equation}
	In the small-field limit, we recover the Weyl transformation $\Omega^2$ in Eq.~\eqref{eq_Weyl_transformation}, and the arrow indicates that we canonically normalized the modulus field.
    
    We compute the modulus-MSSM interactions by plugging in the K\"ahler potential in Eq.~\eqref{eq_SUGRA_app_Kahler_potential} above into the fermion Lagrangian in Eq.~\eqref{eq_SUGRA_fermion_sector}. The fermion field must be canonically normalized as $\Omega\psi\rightarrow\psi$. After this redefinition, we recover the usual coupling of the dilaton to the fermion masses that are generated by the superpotential $W=\frac{1}{2}m\bs\Phi^2$,\begin{equation}
		\L_{\textrm{MSSM-}\chi}\supset-\frac{1}{2}\Omega m\psi\psi+\hc
	\end{equation}
	This is in agreement with the dilaton-electron coupling in Eq.~\eqref{Eq:chi_SM}, which is found by the usual Weyl transformation. The covariant fermion derivative in Eq.~\eqref{eq_SUGRA_covaraint_fermion_derivative} induces a derivative R-axion coupling in addition to the standard kinetic term,
	\begin{equation}
		\frac{1}{\sqrt{-g}}\L_\psi\supset\frac{i}{2}(\psi^{\bar i \dagger}\bar\sigma^\mu \partial_\mu\psi^i-\partial_\mu\psi^{\bar i\dagger}\bar\sigma^{\mu}\psi^i)+\frac{\chi^2}{6M_\pl^2}\partial_\mu\vartheta \psi^{\bar i\dagger}\bar\sigma^{\mu}\psi^i.
	\end{equation}
	After canonical normalization of the R-axion as $\chimin\vartheta\rightarrow\vartheta$, we see that the R-axion couples derivatively to the axial fermion current with the same strength $\Lambda$ as the dilaton couples to $T^\mu_\mu$ [cf. Eq.~\eqref{eq_pheno_dilaton_SM_trace}],
	\begin{equation}\label{eq_SUGRA_R-axion_MSSM_couplings}
		\L_{\textrm{MSSM-}\vartheta}=\frac{1}{\Lambda}\partial_\mu\vartheta J_A^\mu.
	\end{equation}
	By choosing $\psi^j=\tilde\omega$ in Eq.~\eqref{eq_SUGRA_fermion_sector} we obtain the dilatino couplings,
	\begin{equation}\label{eq_SUGRA_dilatino_MSSM_couplings}
		\L_{\textrm{MSSM-}\tilde\omega}\supset-\frac{\omega^*}{3M_\pl^2}W_i\psi^i\tilde\omega+\hc\sim \frac{1}{\Lambda}O(y\tilde q h q\tilde\omega, y \tilde q^2 \tilde h \tilde\omega, \mu h \tilde h \tilde\omega).
	\end{equation}
	Here $y$ stands for the Yukawa couplings, $q$ for quarks and leptons, $\tilde q$ for squarks and sleptons, $h$ for the Higgs scalars, $\tilde h$ for the Higgsino and $\mu$ is the usual $\mu$ parameter.
	
	In summary, the interactions of the MSSM with all components of the modulus supermultiplet are suppressed by the same trans-Planckian scale $\Lambda$ defined in Eq.~\eqref{eq_pheno_dilaton_SM_trace}. The implications of the R-axion and dilatino couplings to the MSSM are discussed in Sec.~\ref{subsec_r-axion_dilatino_pheno} in the main text.
	
	\bibliographystyle{JHEP}
	\bibliography{references}

@article{Randall:1999ee,
    author = "Randall, Lisa and Sundrum, Raman",
    title = "{A Large mass hierarchy from a small extra dimension}",
    eprint = "hep-ph/9905221",
    archivePrefix = "arXiv",
    reportNumber = "MIT-CTP-2860, PUPT-1860, BUHEP-99-9",
    doi = "10.1103/PhysRevLett.83.3370",
    journal = "Phys. Rev. Lett.",
    volume = "83",
    pages = "3370--3373",
    year = "1999"
}

@article{Cacciapaglia:2008bi,
    author = "Cacciapaglia, Giacomo and Marandella, Guido and Terning, John",
    title = "{Dimensions of Supersymmetric Operators from AdS/CFT}",
    eprint = "0802.2946",
    archivePrefix = "arXiv",
    primaryClass = "hep-th",
    doi = "10.1088/1126-6708/2009/06/027",
    journal = "JHEP",
    volume = "06",
    pages = "027",
    year = "2009"
}

@article{Marti:2001iw,
    author = "Marti, Daniel and Pomarol, Alex",
    title = "{Supersymmetric theories with compact extra dimensions in N=1 superfields}",
    eprint = "hep-th/0106256",
    archivePrefix = "arXiv",
    reportNumber = "UAB-FT-518",
    doi = "10.1103/PhysRevD.64.105025",
    journal = "Phys. Rev. D",
    volume = "64",
    pages = "105025",
    year = "2001"
}

@article{Luty:2000ec,
    author = "Luty, Markus A. and Sundrum, Raman",
    title = "{Hierarchy stabilization in warped supersymmetry}",
    eprint = "hep-th/0012158",
    archivePrefix = "arXiv",
    reportNumber = "UMD-PP-00-028, JHU-TIPAC-2000-06",
    doi = "10.1103/PhysRevD.64.065012",
    journal = "Phys. Rev. D",
    volume = "64",
    pages = "065012",
    year = "2001"
}

@article{Csaki:2023pwy,
    author = "Cs\'aki, Csaba and Geller, Michael and Heller-Algazi, Zamir and Ismail, Ameen",
    title = "{Relevant dilaton stabilization}",
    eprint = "2301.10247",
    archivePrefix = "arXiv",
    primaryClass = "hep-ph",
    doi = "10.1007/JHEP06(2023)202",
    journal = "JHEP",
    volume = "06",
    pages = "202",
    year = "2023"
}

@article{Gherghetta:2000qt,
    author = "Gherghetta, Tony and Pomarol, Alex",
    title = "{Bulk fields and supersymmetry in a slice of AdS}",
    eprint = "hep-ph/0003129",
    archivePrefix = "arXiv",
    reportNumber = "CERN-TH-2000-081, UNIL-IPT-00-06",
    doi = "10.1016/S0550-3213(00)00392-8",
    journal = "Nucl. Phys. B",
    volume = "586",
    pages = "141--162",
    year = "2000"
}

@article{Goldberger:1999uk,
    author = "Goldberger, Walter D. and Wise, Mark B.",
    title = "{Modulus stabilization with bulk fields}",
    eprint = "hep-ph/9907447",
    archivePrefix = "arXiv",
    reportNumber = "CALT-68-2232",
    doi = "10.1103/PhysRevLett.83.4922",
    journal = "Phys. Rev. Lett.",
    volume = "83",
    pages = "4922--4925",
    year = "1999"
}

@article{Goldberger:1999un,
    author = "Goldberger, Walter D. and Wise, Mark B.",
    title = "{Phenomenology of a stabilized modulus}",
    eprint = "hep-ph/9911457",
    archivePrefix = "arXiv",
    reportNumber = "CALT-68-2250",
    doi = "10.1016/S0370-2693(00)00099-X",
    journal = "Phys. Lett. B",
    volume = "475",
    pages = "275--279",
    year = "2000"
}

@article{Nelson:1993nf,
    author = "Nelson, Ann E. and Seiberg, Nathan",
    title = "{R symmetry breaking versus supersymmetry breaking}",
    eprint = "hep-ph/9309299",
    archivePrefix = "arXiv",
    reportNumber = "UCSD-PTH-93-27, RU-93-42",
    doi = "10.1016/0550-3213(94)90577-0",
    journal = "Nucl. Phys. B",
    volume = "416",
    pages = "46--62",
    year = "1994"
}

@article{Rattazzi:2000hs,
    author = "Rattazzi, R. and Zaffaroni, A.",
    title = "{Comments on the holographic picture of the Randall-Sundrum model}",
    eprint = "hep-th/0012248",
    archivePrefix = "arXiv",
    reportNumber = "SNS-PH-00-21, BICOCCA-FT-00-25",
    doi = "10.1088/1126-6708/2001/04/021",
    journal = "JHEP",
    volume = "04",
    pages = "021",
    year = "2001"
}

@article{Arkani-Hamed:2000ijo,
    author = "Arkani-Hamed, Nima and Porrati, Massimo and Randall, Lisa",
    title = "{Holography and phenomenology}",
    eprint = "hep-th/0012148",
    archivePrefix = "arXiv",
    reportNumber = "LBL-46886, UCB-PTH-00-32, NYU-TH-00-09-01",
    doi = "10.1088/1126-6708/2001/08/017",
    journal = "JHEP",
    volume = "08",
    pages = "017",
    year = "2001"
}

@article{Maru:2003mq,
    author = "Maru, Nobuhito and Okada, Nobuchika",
    title = "{Supersymmetric radius stabilization in warped extra dimensions}",
    eprint = "hep-th/0312148",
    archivePrefix = "arXiv",
    reportNumber = "RIKEN-TH-17, KEK-TH-931",
    doi = "10.1103/PhysRevD.70.025002",
    journal = "Phys. Rev. D",
    volume = "70",
    pages = "025002",
    year = "2004"
}

@article{Randall:1998uk,
    author = "Randall, Lisa and Sundrum, Raman",
    title = "{Out of this world supersymmetry breaking}",
    eprint = "hep-th/9810155",
    archivePrefix = "arXiv",
    reportNumber = "MIT-CTP-2788, PUPT-1815, BUHEP-98-26",
    doi = "10.1016/S0550-3213(99)00359-4",
    journal = "Nucl. Phys. B",
    volume = "557",
    pages = "79--118",
    year = "1999"
}

@article{Giudice:1998xp,
    author = "Giudice, Gian F. and Luty, Markus A. and Murayama, Hitoshi and Rattazzi, Riccardo",
    title = "{Gaugino mass without singlets}",
    eprint = "hep-ph/9810442",
    archivePrefix = "arXiv",
    reportNumber = "CERN-TH-98-337, LBNL-42419, LBL-42419, UCB-PTH-98-50, UMD-PP-99-037",
    doi = "10.1088/1126-6708/1998/12/027",
    journal = "JHEP",
    volume = "12",
    pages = "027",
    year = "1998"
}

@article{Giudice:1998bp,
    author = "Giudice, G. F. and Rattazzi, R.",
    title = "{Theories with gauge mediated supersymmetry breaking}",
    eprint = "hep-ph/9801271",
    archivePrefix = "arXiv",
    reportNumber = "CERN-TH-97-380",
    doi = "10.1016/S0370-1573(99)00042-3",
    journal = "Phys. Rept.",
    volume = "322",
    pages = "419--499",
    year = "1999"
}

@article{Cacciapaglia:2023syp,
    author = "Cacciapaglia, Giacomo and Deandrea, Aldo and Isnard, Wanda",
    title = "{Hidden supersymmetric dark sectors}",
    eprint = "2304.05431",
    archivePrefix = "arXiv",
    primaryClass = "hep-ph",
    doi = "10.1103/PhysRevD.109.015024",
    journal = "Phys. Rev. D",
    volume = "109",
    number = "1",
    pages = "015024",
    year = "2024"
}

@article{Sundrum:2009gv,
    author = "Sundrum, Raman",
    title = "{SUSY Splits, But Then Returns}",
    eprint = "0909.5430",
    archivePrefix = "arXiv",
    primaryClass = "hep-th",
    doi = "10.1007/JHEP01(2011)062",
    journal = "JHEP",
    volume = "01",
    pages = "062",
    year = "2011"
}

@article{Gubser:1999vj,
	author = "Gubser, Steven S.",
	title = "{AdS / CFT and gravity}",
	eprint = "hep-th/9912001",
	archivePrefix = "arXiv",
	reportNumber = "HUTP-99-A065",
	doi = "10.1103/PhysRevD.63.084017",
	journal = "Phys. Rev. D",
	volume = "63",
	pages = "084017",
	year = "2001"
}

@article{Luty:2002ff,
    author = "Luty, Markus A.",
    title = "{Weak scale supersymmetry without weak scale supergravity}",
    eprint = "hep-th/0205077",
    archivePrefix = "arXiv",
    reportNumber = "UMD-PP-02-046",
    doi = "10.1103/PhysRevLett.89.141801",
    journal = "Phys. Rev. Lett.",
    volume = "89",
    pages = "141801",
    year = "2002"
}

@article{Goh:2003yr,
    author = "Goh, Hock-Seng and Luty, Markus A. and Ng, Siew-Phang",
    title = "{Supersymmetry without supersymmetry}",
    eprint = "hep-th/0309103",
    archivePrefix = "arXiv",
    reportNumber = "UMD-PP-03-70",
    doi = "10.1088/1126-6708/2005/01/040",
    journal = "JHEP",
    volume = "01",
    pages = "040",
    year = "2005"
}

@book{Wess:1992cp,
    author = "Wess, J. and Bagger, J.",
    title = "{Supersymmetry and supergravity}",
    isbn = "978-0-691-02530-8",
    publisher = "Princeton University Press",
    address = "Princeton, NJ, USA",
    year = "1992"
}

@book{RauschdeTraubenberg:2020kol,
    author = "Rausch de Traubenberg, Michel and Valenzuela, Mauricio",
    title = "{A Supergravity Primer}: {From Geometrical Principles to the Final Lagrangian}",
    doi = "10.1142/11557",
    isbn = "978-981-12-1051-8, 978-981-12-1053-2",
    publisher = "WSP",
    address = "Singapur",
    year = "2020"
}

@article{Hubisz:2024hyz,
    author = "Hubisz, Jay and Ironi, Shaked and Perez, Gilad and Rosenfeld, Rogerio",
    title = "{A Note on the Quality of Dilatonic Ultralight Dark Matter}",
    eprint = "2401.08737",
    archivePrefix = "arXiv",
    primaryClass = "hep-ph",
    month = "1",
    year = "2024"
}

@article{Csaki:2000zn,
    author = "Csaki, Csaba and Graesser, Michael L. and Kribs, Graham D.",
    title = "{Radion dynamics and electroweak physics}",
    eprint = "hep-th/0008151",
    archivePrefix = "arXiv",
    reportNumber = "SCIPP-00-27",
    doi = "10.1103/PhysRevD.63.065002",
    journal = "Phys. Rev. D",
    volume = "63",
    pages = "065002",
    year = "2001"
}

@article{Luty:2001jh,
    author = "Luty, Markus A. and Sundrum, Raman",
    title = "{Supersymmetry breaking and composite extra dimensions}",
    eprint = "hep-th/0105137",
    archivePrefix = "arXiv",
    reportNumber = "UMD-PP-01-054, JHU-TIPAC-2001-01",
    doi = "10.1103/PhysRevD.65.066004",
    journal = "Phys. Rev. D",
    volume = "65",
    pages = "066004",
    year = "2002"
}

@article{Luty:2001zv,
    author = "Luty, Markus and Sundrum, Raman",
    title = "{Anomaly mediated supersymmetry breaking in four-dimensions, naturally}",
    eprint = "hep-th/0111231",
    archivePrefix = "arXiv",
    reportNumber = "UMD-PP-02-019, JHU-TIPAC-2001-05",
    doi = "10.1103/PhysRevD.67.045007",
    journal = "Phys. Rev. D",
    volume = "67",
    pages = "045007",
    year = "2003"
}

@article{Schmaltz:2006qs,
    author = "Schmaltz, Martin and Sundrum, Raman",
    title = "{Conformal Sequestering Simplified}",
    eprint = "hep-th/0608051",
    archivePrefix = "arXiv",
    doi = "10.1088/1126-6708/2006/11/011",
    journal = "JHEP",
    volume = "11",
    pages = "011",
    year = "2006"
}

@article{ATLAS:2024lda,
    author = "Aad, Georges and others",
    collaboration = "ATLAS",
    title = "{The quest to discover supersymmetry at the ATLAS experiment}",
    eprint = "2403.02455",
    archivePrefix = "arXiv",
    primaryClass = "hep-ex",
    reportNumber = "CERN-EP-2024-056",
    month = "3",
    year = "2024"
}

@article{Irsic:2017ixq,
	author = "Ir\v{s}i\v{c}, Vid and others",
	title = "{New Constraints on the free-streaming of warm dark matter from intermediate and small scale Lyman-$\alpha$ forest data}",
	eprint = "1702.01764",
	archivePrefix = "arXiv",
	primaryClass = "astro-ph.CO",
	doi = "10.1103/PhysRevD.96.023522",
	journal = "Phys. Rev. D",
	volume = "96",
	number = "2",
	pages = "023522",
	year = "2017"
}

@article{Palanque-Delabrouille:2019iyz,
	author = {Palanque-Delabrouille, Nathalie and Y\`eche, Christophe and Sch\"oneberg, Nils and Lesgourgues, Julien and Walther, Michael and Chabanier, Sol\`ene and Armengaud, Eric},
	title = "{Hints, neutrino bounds and WDM constraints from SDSS DR14 Lyman-$\alpha$ and Planck full-survey data}",
	eprint = "1911.09073",
	archivePrefix = "arXiv",
	primaryClass = "astro-ph.CO",
	doi = "10.1088/1475-7516/2020/04/038",
	journal = "JCAP",
	volume = "04",
	pages = "038",
	year = "2020"
}

@article{Garzilli:2019qki,
	author = "Garzilli, Antonella and Magalich, Andrii and Ruchayskiy, Oleg and Boyarsky, Alexey",
	title = "{How to constrain warm dark matter with the Lyman-$\alpha$ forest}",
	eprint = "1912.09397",
	archivePrefix = "arXiv",
	primaryClass = "astro-ph.CO",
	doi = "10.1093/mnras/stab192",
	journal = "Mon. Not. Roy. Astron. Soc.",
	volume = "502",
	number = "2",
	pages = "2356--2363",
	year = "2021"
}

@article{Yeche:2017upn,
	author = "Y\`eche, Christophe and Palanque-Delabrouille, Nathalie and Baur, Julien and du Mas des Bourboux, H\'elion",
	title = "{Constraints on neutrino masses from Lyman-alpha forest power spectrum with BOSS and XQ-100}",
	eprint = "1702.03314",
	archivePrefix = "arXiv",
	primaryClass = "astro-ph.CO",
	doi = "10.1088/1475-7516/2017/06/047",
	journal = "JCAP",
	volume = "06",
	pages = "047",
	year = "2017"
}

@article{Bellazzini:2013fga,
    author = "Bellazzini, Brando and Csaki, Csaba and Hubisz, Jay and Serra, Javi and Terning, John",
    title = "{A Naturally Light Dilaton and a Small Cosmological Constant}",
    eprint = "1305.3919",
    archivePrefix = "arXiv",
    primaryClass = "hep-th",
    doi = "10.1140/epjc/s10052-014-2790-x",
    journal = "Eur. Phys. J. C",
    volume = "74",
    pages = "2790",
    year = "2014"
}

@article{Creminelli:2001th,
	author = "Creminelli, Paolo and Nicolis, Alberto and Rattazzi, Riccardo",
	title = "{Holography and the electroweak phase transition}",
	eprint = "hep-th/0107141",
	archivePrefix = "arXiv",
	reportNumber = "CERN-TH-2001-189",
	doi = "10.1088/1126-6708/2002/03/051",
	journal = "JHEP",
	volume = "03",
	pages = "051",
	year = "2002"
}

@article{Kuhlen:2012ft,
    author = "Kuhlen, Michael and Vogelsberger, Mark and Angulo, Raul",
    title = "{Numerical Simulations of the Dark Universe: State of the Art and the Next Decade}",
    eprint = "1209.5745",
    archivePrefix = "arXiv",
    primaryClass = "astro-ph.CO",
    doi = "10.1016/j.dark.2012.10.002",
    journal = "Phys. Dark Univ.",
    volume = "1",
    pages = "50--93",
    year = "2012"
}

@article{Abbott:1982af,
    author = "Abbott, L. F. and Sikivie, P.",
    editor = "Srednicki, M. A.",
    title = "{A Cosmological Bound on the Invisible Axion}",
    reportNumber = "PRINT-82-0695 (BRANDEIS)",
    doi = "10.1016/0370-2693(83)90638-X",
    journal = "Phys. Lett. B",
    volume = "120",
    pages = "133--136",
    year = "1983"
}

@article{Dine:1982ah,
    author = "Dine, Michael and Fischler, Willy",
    editor = "Srednicki, M. A.",
    title = "{The Not So Harmless Axion}",
    reportNumber = "UPR-0201T",
    doi = "10.1016/0370-2693(83)90639-1",
    journal = "Phys. Lett. B",
    volume = "120",
    pages = "137--141",
    year = "1983"
}

@article{Preskill:1982cy,
    author = "Preskill, John and Wise, Mark B. and Wilczek, Frank",
    editor = "Srednicki, M. A.",
    title = "{Cosmology of the Invisible Axion}",
    reportNumber = "HUTP-82-A048, NSF-ITP-82-103",
    doi = "10.1016/0370-2693(83)90637-8",
    journal = "Phys. Lett. B",
    volume = "120",
    pages = "127--132",
    year = "1983"
}

@article{Turner:1983he,
    author = "Turner, Michael S.",
    title = "{Coherent Scalar Field Oscillations in an Expanding Universe}",
    reportNumber = "EFI-83-29-CHICAGO",
    doi = "10.1103/PhysRevD.28.1243",
    journal = "Phys. Rev. D",
    volume = "28",
    pages = "1243",
    year = "1983"
}

@article{Damour:2010rp,
    author = "Damour, Thibault and Donoghue, John F.",
    title = "{Equivalence Principle Violations and Couplings of a Light Dilaton}",
    eprint = "1007.2792",
    archivePrefix = "arXiv",
    primaryClass = "gr-qc",
    doi = "10.1103/PhysRevD.82.084033",
    journal = "Phys. Rev. D",
    volume = "82",
    pages = "084033",
    year = "2010"
}

@article{Adelberger:2009zz,
    author = "Adelberger, E. G. and Gundlach, J. H. and Heckel, B. R. and Hoedl, S. and Schlamminger, S.",
    title = "{Torsion balance experiments: A low-energy frontier of particle physics}",
    doi = "10.1016/j.ppnp.2008.08.002",
    journal = "Prog. Part. Nucl. Phys.",
    volume = "62",
    pages = "102--134",
    year = "2009"
}

@article{Murata:2014nra,
    author = "Murata, Jiro and Tanaka, Saki",
    title = "{A review of short-range gravity experiments in the LHC era}",
    eprint = "1408.3588",
    archivePrefix = "arXiv",
    primaryClass = "hep-ex",
    doi = "10.1088/0264-9381/32/3/033001",
    journal = "Class. Quant. Grav.",
    volume = "32",
    number = "3",
    pages = "033001",
    year = "2015"
}

@article{Chen:2014oda,
    author = "Chen, Y. -J. and Tham, W. K. and Krause, D. E. and Lopez, D. and Fischbach, Ephraim and Decca, R. S.",
    title = "{Stronger Limits on Hypothetical Yukawa Interactions in the 30\textendash{}8000 nm Range}",
    eprint = "1410.7267",
    archivePrefix = "arXiv",
    primaryClass = "hep-ex",
    doi = "10.1103/PhysRevLett.116.221102",
    journal = "Phys. Rev. Lett.",
    volume = "116",
    number = "22",
    pages = "221102",
    year = "2016"
}

@article{Lee:2020zjt,
    author = "Lee, J. G. and Adelberger, E. G. and Cook, T. S. and Fleischer, S. M. and Heckel, B. R.",
    title = "{New Test of the Gravitational $1/r^2$ Law at Separations down to 52 $\mu$m}",
    eprint = "2002.11761",
    archivePrefix = "arXiv",
    primaryClass = "hep-ex",
    doi = "10.1103/PhysRevLett.124.101101",
    journal = "Phys. Rev. Lett.",
    volume = "124",
    number = "10",
    pages = "101101",
    year = "2020"
}

@article{Antypas:2022asj,
    author = "Antypas, D. and others",
    title = "{New Horizons: Scalar and Vector Ultralight Dark Matter}",
    eprint = "2203.14915",
    archivePrefix = "arXiv",
    primaryClass = "hep-ex",
    reportNumber = "FERMILAB-PUB-22-262-AD-PPD-T",
    month = "3",
    year = "2022"
}

@article{Geraci:2018fax,
    author = "Geraci, Andrew A. and Bradley, Colin and Gao, Dongfeng and Weinstein, Jonathan and Derevianko, Andrei",
    title = "{Searching for Ultralight Dark Matter with Optical Cavities}",
    eprint = "1808.00540",
    archivePrefix = "arXiv",
    primaryClass = "astro-ph.IM",
    doi = "10.1103/PhysRevLett.123.031304",
    journal = "Phys. Rev. Lett.",
    volume = "123",
    number = "3",
    pages = "031304",
    year = "2019"
}

@article{Manley:2019vxy,
    author = "Manley, Jack and Wilson, Dalziel and Stump, Russell and Grin, Daniel and Singh, Swati",
    title = "{Searching for Scalar Dark Matter with Compact Mechanical Resonators}",
    eprint = "1910.07574",
    archivePrefix = "arXiv",
    primaryClass = "astro-ph.IM",
    doi = "10.1103/PhysRevLett.124.151301",
    journal = "Phys. Rev. Lett.",
    volume = "124",
    number = "15",
    pages = "151301",
    year = "2020"
}

@article{Lorenzo:2016isc,
    author = "Lorenzo, L. A. and Schwab, K. C.",
    title = "{Ultra-High Q Acoustic Resonance in Superfluid$^4$ He}",
    doi = "10.1007/s10909-016-1674-x",
    journal = "J. Low Temp. Phys.",
    volume = "186",
    number = "3-4",
    pages = "233--240",
    year = "2017"
}

@article{Rowan:2000wd,
    author = "Rowan, S. and Cagnoli, G. and Sneddon, P. and Hough, J. and Route, R. and Gustafson, E. K. and Fejer, M. M. and Mitrofanov, V.",
    title = "{Investigation of mechanical loss factors of some candidate materials for the test masses of gravitational wave detectors}",
    doi = "10.1016/S0375-9601(99)00874-9",
    journal = "Phys. Lett. A",
    volume = "265",
    pages = "5--11",
    year = "2000"
}

@phdthesis{pillarcitation,
    author = "Neuhaus, Leonhard",
    title = "Cooling a macroscopic mechanical oscillator close to its quantum ground state",
    school = "Universit\'e Pierre et Marie Curie-Paris VI",
    year = "2016"
}

@article{bawcitation1,
  title = {Observation of Rayleigh Phonon Scattering through Excitation of Extremely High Overtones in Low-Loss Cryogenic Acoustic Cavities for Hybrid Quantum Systems},
  author = {Goryachev, Maxim and Creedon, Daniel L. and Galliou, Serge and Tobar, Michael E.},
  journal = {Phys. Rev. Lett.},
  volume = {111},
  issue = {8},
  pages = {085502},
  numpages = {5},
  year = {2013},
  month = {Aug},
  publisher = {American Physical Society},
  doi = {10.1103/PhysRevLett.111.085502},
  url = {https://link.aps.org/doi/10.1103/PhysRevLett.111.085502}
}

@article{bawcitation2,
    author = {Galliou, Serge and Goryachev, Maxim and Bourquin, Roger and Abb\'e, Philippe and Aubry, Jean Pierre and Tobar, Michael E.},
    title = {Extremely Low Loss Phonon-Trapping Cryogenic Acoustic Cavities for Future Physical Experiments},
    journal = {Scientific Reports},
    volume = {3},
    pages = {2132},
    year = {2013},
    doi = {10.1038/srep02132}
}

@article{bawcitation3,
    author = "Arvanitaki, Asimina and Dimopoulos, Savas and Van Tilburg, Ken",
    title = "{Sound of Dark Matter: Searching for Light Scalars with Resonant-Mass Detectors}",
    eprint = "1508.01798",
    archivePrefix = "arXiv",
    primaryClass = "hep-ph",
    doi = "10.1103/PhysRevLett.116.031102",
    journal = "Phys. Rev. Lett.",
    volume = "116",
    number = "3",
    pages = "031102",
    year = "2016"
}

@article{Marsh:2015xka,
    author = "Marsh, David J. E.",
    title = "{Axion Cosmology}",
    eprint = "1510.07633",
    archivePrefix = "arXiv",
    primaryClass = "astro-ph.CO",
    reportNumber = "KCL-PH-TH-2015-50",
    doi = "10.1016/j.physrep.2016.06.005",
    journal = "Phys. Rept.",
    volume = "643",
    pages = "1--79",
    year = "2016"
}

@misc{AxionLimits,
  author       = {Ciaran O'Hare},
  title        = {cajohare/AxionLimits: AxionLimits},
  month        = jul,
  year         = 2020,
  publisher    = {Zenodo},
  version      = {v1.0},
  doi          = {10.5281/zenodo.3932430},
  howpublished = {\url{https://cajohare.github.io/AxionLimits/}}
}

@article{Kozyryev:2018pcp,
    author = "Kozyryev, Ivan and Lasner, Zack and Doyle, John M.",
    title = "{Enhanced sensitivity to ultralight bosonic dark matter in the spectra of the linear radical SrOH}",
    eprint = "1805.08185",
    archivePrefix = "arXiv",
    primaryClass = "physics.atom-ph",
    doi = "10.1103/PhysRevA.103.043313",
    journal = "Phys. Rev. A",
    volume = "103",
    number = "4",
    pages = "043313",
    year = "2021"
}

@article{Badurina:2021rgt,
    author = "Badurina, Leonardo and Buchmueller, Oliver and Ellis, John and Lewicki, Marek and McCabe, Christopher and Vaskonen, Ville",
    title = "{Prospective sensitivities of atom interferometers to gravitational waves and ultralight
 dark matter}",
    eprint = "2108.02468",
    archivePrefix = "arXiv",
    primaryClass = "gr-qc",
    reportNumber = "AION-REPORT/2021-04, KCL-PH-TH/2021-61, CERN-TH-2021-116",
    doi = "10.1098/rsta.2021.0060",
    journal = "Phil. Trans. A. Math. Phys. Eng. Sci.",
    volume = "380",
    number = "2216",
    pages = "20210060",
    year = "2021"
}

@article{MAGIS-100:2021etm,
    author = "Abe, Mahiro and others",
    collaboration = "MAGIS-100",
    title = "{Matter-wave Atomic Gradiometer Interferometric Sensor (MAGIS-100)}",
    eprint = "2104.02835",
    archivePrefix = "arXiv",
    primaryClass = "physics.atom-ph",
    reportNumber = "FERMILAB-PUB-21-031-AD-DI-FESS-QIS-T",
    doi = "10.1088/2058-9565/abf719",
    journal = "Quantum Sci. Technol.",
    volume = "6",
    number = "4",
    pages = "044003",
    year = "2021"
}

@article{Bellazzini:2014yua,
	author = "Bellazzini, Brando and Cs\'aki, Csaba and Serra, Javi",
	title = "{Composite Higgses}",
	eprint = "1401.2457",
	archivePrefix = "arXiv",
	primaryClass = "hep-ph",
	doi = "10.1140/epjc/s10052-014-2766-x",
	journal = "Eur. Phys. J. C",
	volume = "74",
	number = "5",
	pages = "2766",
	year = "2014"
}

@book{Panico:2015jxa,
	author = "Panico, Giuliano and Wulzer, Andrea",
	title = "{The Composite Nambu-Goldstone Higgs}",
	eprint = "1506.01961",
	archivePrefix = "arXiv",
	primaryClass = "hep-ph",
	reportNumber = "DFPD-2015TH9",
	doi = "10.1007/978-3-319-22617-0",
	publisher = "Springer",
	volume = "913",
	year = "2016"
}

@inbook{Csaki:2018muy,
	author = "Cs\'aki, Csaba and Lombardo, Salvator and Telem, Ofri",
	editor = "Essig, Rouven and Low, Ian",
	title = "{TASI Lectures on Non-supersymmetric BSM Models}",
	booktitle = "{Proceedings,  Theoretical Advanced Study Institute in Elementary Particle Physics : Anticipating the Next Discoveries in Particle Physics (TASI 2016)}: {Boulder, CO, USA, June 6-July 1, 2016}",
	eprint = "1811.04279",
	archivePrefix = "arXiv",
	primaryClass = "hep-ph",
	doi = "10.1142/9789813233348_0007",
	publisher = "WSP",
	pages = "501--570",
	year = "2018"
}

@inproceedings{Csaki:2004ay,
	author = "Csaki, Csaba",
	title = "{TASI lectures on extra dimensions and branes}",
	booktitle = "{Theoretical Advanced Study Institute in Elementary Particle Physics (TASI 2002): Particle Physics and Cosmology: The Quest for Physics Beyond the Standard Model(s)}",
	eprint = "hep-ph/0404096",
	archivePrefix = "arXiv",
	pages = "605--698",
	month = "4",
	year = "2004"
}

@article{Cho:1998js,
    author = "Cho, Y. M. and Keum, Y. Y.",
    title = "{Dilatonic dark matter and unified cosmology: A new paradigm}",
    doi = "10.1088/0264-9381/15/4/013",
    journal = "Class. Quant. Grav.",
    volume = "15",
    pages = "907--921",
    year = "1998"
}

@article{Damour:2010rm,
    author = "Damour, Thibault and Donoghue, John F.",
    title = "{Phenomenology of the Equivalence Principle with Light Scalars}",
    eprint = "1007.2790",
    archivePrefix = "arXiv",
    primaryClass = "gr-qc",
    doi = "10.1088/0264-9381/27/20/202001",
    journal = "Class. Quant. Grav.",
    volume = "27",
    pages = "202001",
    year = "2010"
}

@article{Arvanitaki:2014faa,
    author = "Arvanitaki, Asimina and Huang, Junwu and Van Tilburg, Ken",
    title = "{Searching for dilaton dark matter with atomic clocks}",
    eprint = "1405.2925",
    archivePrefix = "arXiv",
    primaryClass = "hep-ph",
    doi = "10.1103/PhysRevD.91.015015",
    journal = "Phys. Rev. D",
    volume = "91",
    number = "1",
    pages = "015015",
    year = "2015"
}

@article{Planck:2018jri,
    author = "Akrami, Y. and others",
    collaboration = "Planck",
    title = "{Planck 2018 results. X. Constraints on inflation}",
    eprint = "1807.06211",
    archivePrefix = "arXiv",
    primaryClass = "astro-ph.CO",
    doi = "10.1051/0004-6361/201833887",
    journal = "Astron. Astrophys.",
    volume = "641",
    pages = "A10",
    year = "2020"
}

@article{Hook:2018dlk,
    author = "Hook, Anson",
    title = "{TASI Lectures on the Strong CP Problem and Axions}",
    eprint = "1812.02669",
    archivePrefix = "arXiv",
    primaryClass = "hep-ph",
    journal = "PoS",
    volume = "TASI2018",
    pages = "004",
    year = "2019"
}

@article{Turner:1985si,
    author = "Turner, Michael S.",
    title = "{Cosmic and Local Mass Density of Invisible Axions}",
    reportNumber = "FERMILAB-PUB-85-149-A, EFI-85-67-CHICAGO, FERMILAB-PUB-85-093-A",
    doi = "10.1103/PhysRevD.33.889",
    journal = "Phys. Rev. D",
    volume = "33",
    pages = "889--896",
    year = "1986"
}

@article{Lyth:1991ub,
    author = "Lyth, D. H.",
    title = "{Axions and inflation: Sitting in the vacuum}",
    reportNumber = "LANC-TH-91-02-REV-2, LANC-TH-91-02",
    doi = "10.1103/PhysRevD.45.3394",
    journal = "Phys. Rev. D",
    volume = "45",
    pages = "3394--3404",
    year = "1992"
}

@article{Strobl:1994wk,
    author = "Strobl, Karl and Weiler, Thomas J.",
    title = "{Anharmonic evolution of the cosmic axion density spectrum}",
    eprint = "astro-ph/9405028",
    archivePrefix = "arXiv",
    reportNumber = "DAMTP-94-21, VAND-TH-94-4",
    doi = "10.1103/PhysRevD.50.7690",
    journal = "Phys. Rev. D",
    volume = "50",
    pages = "7690--7702",
    year = "1994"
}

@article{Bae:2008ue,
    author = "Bae, Kyu Jung and Huh, Ji-Haeng and Kim, Jihn E.",
    title = "{Update of axion CDM energy}",
    eprint = "0806.0497",
    archivePrefix = "arXiv",
    primaryClass = "hep-ph",
    doi = "10.1088/1475-7516/2008/09/005",
    journal = "JCAP",
    volume = "09",
    pages = "005",
    year = "2008"
}

@article{Visinelli:2009zm,
    author = "Visinelli, Luca and Gondolo, Paolo",
    title = "{Dark Matter Axions Revisited}",
    eprint = "0903.4377",
    archivePrefix = "arXiv",
    primaryClass = "astro-ph.CO",
    doi = "10.1103/PhysRevD.80.035024",
    journal = "Phys. Rev. D",
    volume = "80",
    pages = "035024",
    year = "2009"
}

@article{Chatrchyan:2023cmz,
    author = {Chatrchyan, Aleksandr and Er\"oncel, Cem and Koschnitzke, Matthias and Servant, G\'eraldine},
    title = "{ALP dark matter with non-periodic potentials: parametric resonance, halo formation and gravitational signatures}",
    eprint = "2305.03756",
    archivePrefix = "arXiv",
    primaryClass = "hep-ph",
    reportNumber = "DESY-23-060",
    doi = "10.1088/1475-7516/2023/10/068",
    journal = "JCAP",
    volume = "10",
    pages = "068",
    year = "2023"
}

@article{Kobayashi:2013nva,
    author = "Kobayashi, Takeshi and Kurematsu, Ryosuke and Takahashi, Fuminobu",
    title = "{Isocurvature Constraints and Anharmonic Effects on QCD Axion Dark Matter}",
    eprint = "1304.0922",
    archivePrefix = "arXiv",
    primaryClass = "hep-ph",
    reportNumber = "TU-933",
    doi = "10.1088/1475-7516/2013/09/032",
    journal = "JCAP",
    volume = "09",
    pages = "032",
    year = "2013"
}

@article{Schmitz:2018nhb,
    author = "Schmitz, Kai and Yanagida, Tsutomu T.",
    title = "{Axion Isocurvature Perturbations in Low-Scale Models of Hybrid Inflation}",
    eprint = "1806.06056",
    archivePrefix = "arXiv",
    primaryClass = "hep-ph",
    reportNumber = "IPMU 18-0118",
    doi = "10.1103/PhysRevD.98.075003",
    journal = "Phys. Rev. D",
    volume = "98",
    number = "7",
    pages = "075003",
    year = "2018"
}

@article{Jaeckel:2022kwg,
    author = "Jaeckel, Joerg and Rybka, Gray and Winslow, Lindley",
    title = "{Report of the Topical Group on Wave Dark Matter for Snowmass 2021}",
    eprint = "2209.08125",
    archivePrefix = "arXiv",
    primaryClass = "hep-ph",
    month = "9",
    year = "2022"
}

@article{Hui:2016ltb,
    author = "Hui, Lam and Ostriker, Jeremiah P. and Tremaine, Scott and Witten, Edward",
    title = "{Ultralight scalars as cosmological dark matter}",
    eprint = "1610.08297",
    archivePrefix = "arXiv",
    primaryClass = "astro-ph.CO",
    doi = "10.1103/PhysRevD.95.043541",
    journal = "Phys. Rev. D",
    volume = "95",
    number = "4",
    pages = "043541",
    year = "2017"
}

@article{Hui:2021tkt,
    author = "Hui, Lam",
    title = "{Wave Dark Matter}",
    eprint = "2101.11735",
    archivePrefix = "arXiv",
    primaryClass = "astro-ph.CO",
    doi = "10.1146/annurev-astro-120920-010024",
    journal = "Ann. Rev. Astron. Astrophys.",
    volume = "59",
    pages = "247--289",
    year = "2021"
}

@article{Broadberry:2024yjw,
    author = "Broadberry, Edward and Das, Saurav and Hook, Anson and Marques Tavares, Gustavo",
    title = "{Adiabatic conversion of ALPs into dark photon dark matter}",
    eprint = "2408.03370",
    archivePrefix = "arXiv",
    primaryClass = "hep-ph",
    doi = "10.1007/JHEP03(2025)215",
    journal = "JHEP",
    volume = "03",
    pages = "215",
    year = "2025"
}

@article{Piazza:2010ye,
    author = "Piazza, Federico and Pospelov, Maxim",
    title = "{Sub-eV scalar dark matter through the super-renormalizable Higgs portal}",
    eprint = "1003.2313",
    archivePrefix = "arXiv",
    primaryClass = "hep-ph",
    doi = "10.1103/PhysRevD.82.043533",
    journal = "Phys. Rev. D",
    volume = "82",
    pages = "043533",
    year = "2010"
}

@article{Banerjee:2018xmn,
    author = "Banerjee, Abhishek and Kim, Hyungjin and Perez, Gilad",
    title = "{Coherent relaxion dark matter}",
    eprint = "1810.01889",
    archivePrefix = "arXiv",
    primaryClass = "hep-ph",
    doi = "10.1103/PhysRevD.100.115026",
    journal = "Phys. Rev. D",
    volume = "100",
    number = "11",
    pages = "115026",
    year = "2019"
}

@article{Brzeminski:2020uhm,
    author = "Brzeminski, Dawid and Chacko, Zackaria and Dev, Abhish and Hook, Anson",
    title = "{Time-varying fine structure constant from naturally ultralight dark matter}",
    eprint = "2012.02787",
    archivePrefix = "arXiv",
    primaryClass = "hep-ph",
    doi = "10.1103/PhysRevD.104.075019",
    journal = "Phys. Rev. D",
    volume = "104",
    number = "7",
    pages = "075019",
    year = "2021"
}

@article{Chacko:2012sy,
    author = "Chacko, Zackaria and Mishra, Rashmish K.",
    title = "{Effective Theory of a Light Dilaton}",
    eprint = "1209.3022",
    archivePrefix = "arXiv",
    primaryClass = "hep-ph",
    doi = "10.1103/PhysRevD.87.115006",
    journal = "Phys. Rev. D",
    volume = "87",
    number = "11",
    pages = "115006",
    year = "2013"
}

@article{Coradeschi:2013gda,
    author = "Coradeschi, Francesco and Lodone, Paolo and Pappadopulo, Duccio and Rattazzi, Riccardo and Vitale, Lorenzo",
    title = "{A naturally light dilaton}",
    eprint = "1306.4601",
    archivePrefix = "arXiv",
    primaryClass = "hep-th",
    doi = "10.1007/JHEP11(2013)057",
    journal = "JHEP",
    volume = "11",
    pages = "057",
    year = "2013"
}

@article{Girmohanta:2024ywz,
    author = "Girmohanta, Sudhakantha and Nakai, Yuichiro and Qiu, Yu-Cheng and Zhang, Zhihao",
    title = "{Wiggly dilaton: A landscape of spontaneously broken scale invariance}",
    eprint = "2411.16304",
    archivePrefix = "arXiv",
    primaryClass = "hep-th",
    doi = "10.1016/j.physletb.2025.139534",
    journal = "Phys. Lett. B",
    volume = "866",
    pages = "139534",
    year = "2025"
}

@article{Cyncynates:2024bxw,
    author = "Cyncynates, David and Simon, Olivier",
    title = "{Minimal targets for dilaton direct detection}",
    eprint = "2408.16816",
    archivePrefix = "arXiv",
    primaryClass = "hep-ph",
    doi = "10.1103/8mc9-6cmr",
    journal = "Phys. Rev. D",
    volume = "112",
    number = "5",
    pages = "055002",
    year = "2025"
}

@article{Cyncynates:2024ufu,
    author = "Cyncynates, David and Simon, Olivier",
    title = "{Scalar Relics from the Hot Big Bang}",
    eprint = "2410.22409",
    archivePrefix = "arXiv",
    primaryClass = "hep-ph",
    doi = "10.1103/wjt8-9dh9",
    journal = "Phys. Rev. Lett.",
    volume = "135",
    number = "10",
    pages = "101003",
    year = "2025"
}
	
\end{document}